\documentclass[preprint,11pt]{elsarticle}
\usepackage[latin1]{inputenc}
\usepackage[english]{babel}
\usepackage{amssymb,amsmath,graphicx}
\usepackage{multirow}
\usepackage{enumitem}
\usepackage{setspace}
\usepackage[table]{xcolor}
\usepackage{microtype}
\usepackage{varwidth}
\usepackage{afterpage,lscape}
\newcolumntype{M}[1]{>{\centering\let\newline\\\arraybackslash\hspace{0pt}}m{#1}}
\usepackage{hhline}
\usepackage{nicefrac}
\usepackage[norelsize,boxed,linesnumbered,noend]{algorithm2e}
\SetArgSty{textrm} 
\SetAlCapSkip{0.7em} 
\usepackage[hidelinks]{hyperref}
\usepackage{multicol}
\usepackage[FIGTOPCAP]{subfigure}
\usepackage{caption}
\usepackage{natbib}
\usepackage{siunitx} 

\usepackage{xfrac}
\usepackage{bbm}

\newcolumntype{P}[1]{>{\centering}p{#1}}
\newcolumntype{H}{>{\setbox0=\hbox\bgroup}c<{\egroup}@{}}
\definecolor{light-gray}{gray}{0.95}

\DeclareGraphicsExtensions{.pdf, .png}

\allowdisplaybreaks
\title{Community Detection in the Stochastic Block Model by Mixed Integer Programming}

\author{Breno Serrano, Thibaut Vidal}

\begin{document}

\linespread{1.6}\selectfont

\begin{center}

\vspace*{-0.5cm}

\begin{huge}
Community Detection in the Stochastic Block Model \\
by Mixed Integer Programming
\end{huge}

\vspace*{0.6cm}

\textbf{Breno Serrano$^{a*}$, Thibaut Vidal$^{a}$} \\
$^a$ Departamento de Inform\'{a}tica, Pontif\'{i}cia Universidade Cat\'{o}lica do Rio de Janeiro (PUC-Rio),  Rua Marqu\^{e}s de S\~{a}o Vicente, 225 - G\'{a}vea, Rio de Janeiro - RJ, 22451-900, Brazil \\
\{vidalt,bserrano\}@inf.puc-rio.br \\
\vspace*{0.15cm}

\end{center}
\noindent
\textbf{Abstract.}
The Degree-Corrected Stochastic Block Model (DCSBM) is a popular model to generate random graphs with community structure given an expected degree sequence.
The standard approach of community detection based on the DCSBM is to search for the model parameters that are the most likely to have produced the observed network data through maximum likelihood estimation (MLE).
Current techniques for the MLE problem are heuristics, and therefore do not guarantee convergence to the optimum. We present mathematical programming formulations and exact solution methods that can provably find the model parameters and community assignments of maximum likelihood given an observed graph. We compare these exact methods with classical heuristic algorithms based on expectation-maximization (EM). The solutions given by exact methods give us a principled way of measuring the experimental performance of classical heuristics and comparing different variations thereof.
\vspace*{0.2cm}

\vspace*{-0.5cm}

\noindent
$^*$ Corresponding author

\noindent
Declarations of interest: none

\linespread{2.0}\selectfont

\section{Introduction}
\label{Intro}

In the community detection problem, we observe a graph $G=(V,E)$ and aim to find groups of vertices (or \textit{communities}) which present a similar connection pattern \cite{fortunato2016community}.
Some important applications of community detection include the study of social networks \cite{goldenberg2010survey,qi2014optimal,zhao2011new} and predicting the functional family of proteins~\cite{mallek2015community}, among others. One of the most popular approaches for this task consists in fitting a generative model (such as the DCSBM) to the observed graph $G$, and to search for the parameters which maximize the likelihood of the model.

In the DCSBM, the number of edges connecting any two vertices $i$ and $j$ only depends on their group memberships $g_i$ and $g_j$ and on the set of parameters $\theta_i$ which control the expected degree of each vertex $i$.
The DCSBM is characterized by a $K \times K$ affinity matrix $\boldsymbol{\Omega} = (\omega_{rs})$, where $K$ is the number of communities in the graph.
The number of edges between any two nodes $i$ and $j$ is drawn from a Poisson distribution with mean $\theta_i \theta_j \omega_{g_i g_j}$. 
The probability that the observed network $G$, represented by the adjacency matrix $A$, was generated from the DCSBM can be expressed as:
\begin{equation}\label{probability-graph}
\begin{aligned}
P(A | \boldsymbol{g}, \boldsymbol{\Omega}, \boldsymbol{\theta}) = & \prod_{i<j} \frac{(\theta_i \theta_j \omega_{g_{i} g_{j}})^{A_{i j}}}{A_{i j} !} \exp{(-\theta_i \theta_j \omega_{g_{i} g_{j}})} \times \\
& \prod_{i} \frac{\left(\frac{1}{2} \theta_i^2 \omega_{g_{i} g_{i}}\right)^{A_{ii} / 2}}{\left(\frac{1}{2} A_{i i}\right) !} \exp{\left(-\tfrac{1}{2} \theta_i^2 \omega_{g_{i} g_{i}}\right)}
\end{aligned}
\end{equation}
which defines the likelihood function of the DCSBM.
As in \citet{newman2016sbm_modularity}, we consider in this work the case where $\theta_i \theta_j = \frac{k_i k_j}{2m}$, where $\frac{k_i k_j}{2m}$ corresponds to the expected number of edges in the configuration model.
After applying the log on both sides and grouping together constant terms, the log-likelihood function becomes:
\begin{equation}
\label{log-likelihood-dc-sbm}
\log P(A | \boldsymbol{g}, \boldsymbol{\Omega}) = \tfrac{1}{2} \sum_{i,j}^{n} \left(A_{ij} \log{\omega_{g_i g_j}} - \tfrac{k_i k_j}{2 m} \omega_{g_i g_j} \right) + \text{Const}.
\end{equation}
The MLE problem consists in finding the affinity matrix $\boldsymbol{\Omega}^{\text{ML}}$ and group membership assignments $\boldsymbol{g}^{\text{ML}}$ that maximize the log-likelihood function (\ref{log-likelihood-dc-sbm}).

The most common methods for this problem are heuristics, which are not guaranteed to converge to the solution of maximum likelihood.
Theoretical convergence guarantees typically focus on the probability of recovering the \textit{true} underlying communities of a graph generated by the DCSBM in the asymptotic limit where the size of the network grows to infinity. 
Most studies (see, e.g., \cite{abbe2017community}) adopt a statistical and information-theoretic viewpoint and provide thresholds and conditions under which different types of algorithms recover the underlying community assignments with high probability for different asymptotic regimes. 
In this work, instead, we adopt a combinatorial optimization viewpoint, proposing mixed integer programming (MIP) formulations and exact solution methods that can provably find the optimal solution of the maximum likelihood model for an observed graph.

Firstly, we propose a simple descriptive formulation, which results in a mixed integer non-linear program (MINLP). 
This model can be solved to optimality with algorithms based on spatial branch-and-bound (sBB).
Building upon this first formulation, we employ linearization techniques to produce a mixed integer linear programming (MILP) formulation. To improve the formulation, we rely on a dynamic generation of valid inequalities along with symmetry-breaking constraints. Moreover, we carefully analyze the problem to derive tight bounds on the model variables, permitting significant reductions of computational effort to find optimal solutions.

Our solution approach is motivated by recent advances in the field of mathematical programming. Due to extensive research, optimization solvers have continuously improved, permitting the solution of increasingly larger problems. Even though MILP is $\mathcal{NP}$-hard in general, it is now possible to solve instances of dimensions that were absolutely out of reach of solution methods a few decades ago. This increase in computational efficiency is a consequence of methodological improvements which, coupled with hardware improvements, have resulted in speed-up factors of the order of $10^{11}$ over two decades (based on the figures reported in~\cite{bixby2012brief}). 

The study of exact solution algorithms is also an important methodological step for machine learning research.
In particular, exact algorithms provide benchmark solutions that can be used to evaluate the performance of heuristics, measuring how far they are from the optima of the maximum likelihood model. To initiate such an analysis, we consider three basic variants of the EM algorithm, and run computational experiments to compare the heuristic solutions with those found by the exact solution approaches in terms of likelihood and proximity to the ground truth.

Therefore, this work makes a significant step towards developing exact solution methods for the problem of MLE of the DCSBM. Specifically, our contribution is threefold. First, we propose a MILP formulation and an optimal solution algorithm for the problem. To the best of our knowledge, this is the first work to investigate exact methods for the DCSBM.
Second, we perform computational experiments on synthetic graphs to assess the performance of the exact methods, showing that the algorithm based on the MILP formulation clearly outperforms that of the MINLP formulation.
Third, we compare the exact methods with some classical heuristics based on the EM algorithm. These experiments show that MIP approaches are an important asset to evaluate the performance of heuristics and highlight the importance of optimal algorithms in machine learning research. All data sets and source codes needed to reproduce our results are available at: \url{github.com/vidalt/Optimal-SBM}.

The remainder of this paper is structured as follows.
Section \ref{section-related-works} draws an overview of related works. Section \ref{section-exact} introduces our mathematical programming formulations and solution techniques.
Section \ref{section-heuristic} describes the heuristic approaches and Section \ref{section-experiments} presents the computational experiments comparing the proposed solution methods.
Finally, Section \ref{section-conclusions} concludes. 

\section{Related Works}
\label{section-related-works}

A large body of literature on SBMs focuses on community recovery from an information-theoretical and statistical viewpoint.
Special cases of the SBM have regularly been considered, such as the (balanced) Planted Partition Model (PPM). We refer the reader to the survey by \citet{abbe2017community} for a detailed description of different recovery requirements and consistency analyses of various algorithms.

Some studies have explored exact solution methods for community detection based on modularity maximization \cite{aloise2010modularity, xu2007finding}, which is known to be $\mathcal{NP}$-hard \cite{brandes2007modularity}. 
\citet{newman2016sbm_modularity} shows that modularity maximization is equivalent to MLE of the PPM.
The problem of modularity maximization is directly formulated as a mixed integer quadratic program (MIQP) by \citet{xu2007finding} and solved using a branch-and-bound method. 
The authors discuss the use of symmetry-breaking constraints to improve the efficiency of the branch-and-bound exploration. 
\citet{aloise2010modularity} employ techniques based on column generation to improve on previous works \cite{xu2007finding}, reporting a reduction in computing time, and solving larger instances with up to 512 vertices to optimality (vs. 105 vertices in previous works). Modularity maximization (equivalently MLE of the PPM) is, however, a very specialized case of community detection, and no exact solution algorithm has been proposed for MLE of the general SBM to date.

Several studies proposed algorithms for community recovery based on SDP relaxations of the MLE model~\cite{cai2015robust, chen2012clustering, chen2016statistical}, leading to new results regarding the recovery of communities in SBMs with general~$K$  under an implicit assumption of strong assortativity.
\citet{amini2018semidefinite} proposed an SDP relaxation that is tighter than previous ones and works for a broader class of SBMs, including for disassortative structures.

\citet{pia2020linear} considered the problem of exact community recovery for the assortative planted bisection model and discussed the theoretical performance of linear programming (LP) relaxations of the minimum bisection problem for community recovery. They derived sufficient and necessary conditions for recovery using the LP relaxation for different asymptotic regimes.

Algorithms based on expectation-maximization for MLE have been investigated, for example, by~\cite{snijders1997estimation} for the SBM with two communities. 
However, their method is practical only for small graphs.
For large graphs, they introduce a Bayesian estimation method based on Gibbs sampling.
The EM algorithm is also used to maximize the pseudo-likelihood of the SBM parameters in  \citet{amini2013pseudo}.
The general idea of pseudo-likelihood is to approximate the likelihood by ignoring some of the dependency structure of the data to make the model more tractable.

Metaheuristics have been applied to various related clustering problems. Among others, \citet{gribel2019hg} proposed a hybrid genetic algorithm for the minimum sum-of-squares clustering problem, and \citet{hansen2012vns} proposed a variable neighborhood search heuristic for normalized cut clustering.

More broadly, several related works investigated the application of mixed integer programming to classical machine learning models.
Mixed integer optimization has been used to learn optimal decision trees~\cite{bertsimas2017optimal}, Gaussian mixture models (GMM)~\cite{bandi2019learning} and ramp-loss SVMs~\cite{belotti2016indconstr}, among others.
We refer to \cite{bennett2006interplay, gambella2019optimization} for surveys on mathematical programming applied to popular machine learning models.

\section{Solving the DCSBM to Optimality}
\label{section-exact}

This section introduces mathematical programming formulations for the MLE problem given by Equation~(\ref{log-likelihood-dc-sbm}). 
We first present a descriptive formulation as a MINLP model.
Then, we employ different techniques to linearize the model, leading to a MILP formulation. To further improve computational efficiency, we discuss the use of bounds-tightening and symmetry-breaking techniques.

\subsection{Formulation as a Mixed Integer Non-Linear Program}
\label{desc-formulation}

Let $z_{ir}$ be a binary variable which takes value $1$ if vertex $i \in V$ is assigned to community $r \in \mathcal{C}$ and $0$ otherwise, where $\mathcal{C} = \{1, \dots, K\}$ represent the possible communities.
The continuous variables~$\omega_{rs}$, for $r,s \in \mathcal{C}$, represent the elements of the connectivity matrix $\boldsymbol{\Omega}$.
The MLE problem (\ref{log-likelihood-dc-sbm}) can be modeled as the following MINLP, where we minimize the negative log-likelihood (the constant term was omitted):
\begin{align}\label{descriptive-formulation-start}
& \underset{\mathbf{Z},\boldsymbol{\Omega}}{\text{minimize}}
& & \dfrac{1}{2} \sum_{i,j}^{n} \sum_{r,s}^{K} f_{ij}(\omega_{rs}) \, z_{ir} z_{js} \\
& \text{subject to}
& & \sum_{r=1}^{K} z_{ir} = 1 && \forall i \in V \label{assignment-const} \\
& & & z_{ir} \in \{0, 1\} && \forall i \in V, \, r \in \mathcal{C}\\
& & & \omega_{rs} \in \mathbb{R^+} && \forall r,s \in \mathcal{C}\label{descriptive-formulation-end}
\end{align}
where
\begin{equation}
f_{ij}(\omega_{rs}) = -A_{ij} \log{\omega_{rs}} + \tfrac{k_i k_j}{2m} \omega_{rs}.
\end{equation}
In this model, Constraints (\ref{assignment-const}) ensure that each vertex is assigned to exactly one community. This model can be solved to optimality by global optimization solvers such as \texttt{Couenne} \cite{belotti2009couenne}.
However, solution time quickly increases with the size of the networks. In the next sections, we propose some techniques to linearize Model~(\ref{descriptive-formulation-start})--(\ref{descriptive-formulation-end}) into a MILP along with additional refinements that permit a significant reduction of solution time in comparison to the MINLP.

\subsection{Formulation as a Mixed Integer Linear Program}\label{milp-section}

The linearization of the MINLP formulation is done in several steps. First of all, we linearize the function $f_{ij}(\omega_{rs})$ by piecewise outer-approximation.
The function $f_{ij}$ is convex everywhere in its domain, for $A_{ij} > 0$, since:
$\frac{\partial^2 f_{ij}}{\partial \omega^2} = \frac{A_{ij}}{\omega^2} > 0$, $\forall \omega \in \mathbb{R}^+$.
Thus, the value of $f_{ij}$ is always greater than or equal to its tangent calculated at any point $\widetilde{\omega} \in \mathbb{R}^+$:
\begin{equation}
    f_{ij}(\omega) \geq a_{ij\tilde{\omega}} \, \omega + b_{ij\tilde{\omega}} \qquad \forall \omega \in \mathbb{R}^+,
\end{equation}
where the coefficients $a_{ij\tilde{\omega}}$ and $b_{ij\tilde{\omega}}$ defining the tangent line are given by:
\begin{equation}
a_{ij\tilde{\omega}} = -\dfrac{A_{ij}}{\widetilde{\omega}} + \dfrac{k_i k_j}{2m} 
\end{equation}
\begin{equation}
b_{ij\tilde{\omega}} = f_{ij}(\widetilde{\omega}) - a_{ij\tilde{\omega}} \, \widetilde{\omega} = A_{ij} (1 - \log \widetilde{\omega}).
\end{equation}
By making use of this property, we introduce variables $f_{ijrs}$ to represent the value of $f_{ij}(\omega_{rs})$, and reformulate Model~(\ref{descriptive-formulation-start})--(\ref{descriptive-formulation-end}) into:
\begin{align}\label{opwla-formulation}
& \underset{\mathbf{Z},\boldsymbol{\Omega}, \mathbf{F}}{\text{minimize}}
& & \dfrac{1}{2} \sum_{i,j}^{n} \sum_{r,s}^{K}  f_{ijrs} \, z_{ir} z_{js} \\
& \text{subject to}
& & \sum_{r=1}^{K} z_{ir} = 1 &&\forall i \in V \\
& & & f_{ijrs} \geq a_{ij\tilde{\omega}} \, \omega_{rs} + b_{ij\tilde{\omega}} && \forall i,j \in V, \, \forall r,s \in \mathcal{C}, \, \forall \widetilde{\omega} \in \mathbb{R^+} \label{pwl-constraint} \\
& & & z_{ir} \in \{0, 1\} && \forall i \in V, \, \forall r \in \mathcal{C}\\
& & & \omega_{rs} \in \mathbb{R^+} && \forall r,s \in \mathcal{C} \\
& & & f_{ijrs} \in \mathbb{R} && \forall i,j \in V, \, \forall r,s \in \mathcal{C}. \label{opwla-formulation-end}
\end{align}
This model contains an infinite number of constraints of type (\ref{pwl-constraint}), one for every $\widetilde{\omega} \in \mathbb{R}^+$.

Next, we linearize the objective function.
Let $y_{ijrs}$ denote the product of the binary variables $z_{ir}$ and $z_{js}$ in the objective function ($y_{ijrs} = z_{ir} z_{js}$).
The product of two binary variables can be expressed as a set of linear constraints:
\begin{align}
z_{ir} - y_{ijrs} \geq 0, \\
z_{js} - y_{ijrs} \geq 0, \\
1 - z_{ir} - z_{js} + y_{ijrs} \geq 0.
\end{align}
As a result, the objective function can be expressed as $f_{ijrs} \, y_{ijrs}$, which is a product of a continuous and a binary variable.

To linearize the expression $f_{ijrs} \, y_{ijrs}\,$, we introduce continuous variables $x_{ijrs} = f_{ijrs} \, y_{ijrs} = f_{ijrs} \, z_{ir} z_{js}$.
The non-linear expression $f_{ijrs} \, y_{ijrs}$ can be linearized with the big-M technique by introducing additional constraints:
\begin{align}
x_{ijrs} &\leq \overline{M} \, y_{ijrs},  \\
x_{ijrs} &\geq \underline{M} \, y_{ijrs}, \\
x_{ijrs} &\geq f_{ijrs} - \overline{M} (1 - y_{ijrs}) \label{bigM-end}.
\end{align}
Constraints (\ref{bigM-end}) can be combined with Constraints (\ref{pwl-constraint}), yielding:
\begin{equation}
x_{ijrs} \geq a_{ij\tilde{\omega}} \, \omega_{rs} + b_{ij\tilde{\omega}} - \overline{M} (1 - y_{ijrs}) \qquad \forall \widetilde{\omega} \in \mathbb{R}^+. \label{pwl-constraint-big-M}\\
\end{equation}
It is well known that formulations with big-M constants suffer from a weak continuous relaxation (and therefore from larger solution times) if the lower and upper bounds ($\underline{M}$ and $\overline{M}$) are large in absolute value~\cite{belotti2016indconstr, bonami2015mathematical}.
Section~\ref{bounds-tightening-section} proposes some natural values for these bounds.
The resulting model is a MILP, and it can be solved by conventional branch-and-cut algorithms.

\subsection{Dynamic Constraints Generation}

As mentioned previously, the MILP model has an infinite number of constraints of type (\ref{pwl-constraint-big-M}).
To solve it in practice, we initially only consider a small set of these constraints, for a set of break-points $\widetilde{\omega}_p$ indexed by $p \in \mathcal{B}$. Then, new constraints are dynamically introduced in the model during the solution process. Whenever an integer-feasible solution is found during the branch-and-bound, the algorithm checks if any constraint given by (\ref{pwl-constraint-big-M}) is violated with a tolerance of $\epsilon$.
In this case, the solution is declared infeasible and the violated constraints are added to the model. In effect, the method iteratively refines the approximation of the function $f_{ij}$ until the desired precision of $\epsilon$ is achieved.

\subsection{Bounds Tightening} \label{bounds-tightening-section}

Given a fixed assignment of vertices to communities, the optimal value of $\omega_{rs}$ can be found by solving a convex minimization problem by differentiation. 
We show that $\omega_{rs}$ is bounded above by $2 m \rho$:
\begin{equation}
\label{opt-omega-bound}
\omega_{rs}^* = 2m \left( \dfrac{\sum_{i,j} A_{ij} z_{ir} z_{js}
}{\sum_{i,j} k_i k_j z_{ir} z_{js} 
} \right) \leq 2m \rho,
\end{equation}
where 
\begin{equation}
\label{def-rho}
\rho := \max_{i,j} \Big\{\frac{A_{ij}}{ k_i k_j }\Big\}.
\end{equation}
To see that~(\ref{opt-omega-bound}) holds, observe that it is equivalent to:
\begin{equation}
\sum_{i,j} \left( \tfrac{1}{\rho} A_{ij} - k_i k_j \right) z_{ir} z_{js} \leq 0 
\end{equation}
which is satisfied since:
\begin{equation}
\rho \geq \frac{A_{ij}}{ k_i k_j } \qquad \forall i,j \in V
\end{equation}
by definition of $\rho$ in Equation~\ref{def-rho}.

Let $\omega_{rs}^L$ and $\omega_{rs}^U$ denote the lower and upper bounds, respectively, on $\omega_{rs}$. 
We rely on these bounds to derive bounds $\underline{M_{ijrs}} \leq f_{ijrs} \leq \overline{M_{ijrs}}$.
Recall that $f_{ij}(\omega_{rs}) = -A_{ij} \log{\omega_{rs}} + \tfrac{k_i k_j}{2m} \omega_{rs}$.

If $A_{ij} = 0$, then the expression simplifies to:
\begin{equation}
f_{ij}(\omega_{rs}) = \tfrac{k_i k_j}{2m} \omega_{rs}   
\end{equation}
and therefore $f_{ijrs}$ can be bounded by
$0 \leq f_{ijrs} \leq \tfrac{k_i k_j}{2m} \omega_{rs}^U$.

Otherwise, if $A_{ij} \neq 0$, a lower bound can be obtained by calculating the global minimum of $f_{ij}(\omega_{rs})$ with respect to $\omega_{rs}$.
The minimum can be found by solving
\begin{equation}
\frac{\partial f_{ij}}{\partial \omega_{rs}} = - \dfrac{A_{ij}}{\omega_{rs}} + \dfrac{k_i k_j}{2m} = 0
\end{equation}
implying
\begin{equation}
\hat{\omega}_{rs} = \dfrac{2 m A_{ij}}{k_i k_j}
\end{equation}
and therefore
\begin{equation}\label{lb-M}
\underline{M_{ijrs}} = - A_{ij} \log{\tfrac{2 m A_{ij}}{k_i k_j}} + A_{ij} = A_{ij} \Big( 1 - \log{A_{ij}} + \log{\tfrac{k_i k_j}{2m}} \Big).
\end{equation}
Since $f_{ij}(\omega_{rs})$ is convex, the upper bound $\overline{M_{ijrs}}$ can be defined by calculating the function value at the extreme points of the domain $[\omega_{rs}^L, \omega_{rs}^U]$:
\begin{equation}
\overline{M_{ijrs}} = \max \{f_{ij}(\omega_{rs}^L),f_{ij}(\omega_{rs}^U)\}.
\end{equation}

Overall, the upper and lower bounds are given by:
\begin{gather}\label{upper-bounds}
\overline{M_{ijrs}} = 
\left\{
    \begin{array}{ll}
        \tfrac{k_i k_j}{2m} \omega_{rs}^U,  & \mbox{if } A_{ij} = 0 \\
        \max \{f_{ij}(\omega_{rs}^L),f_{ij}(\omega_{rs}^U)\}, & \mbox{if } A_{ij} \neq 0 
    \end{array}
\right. 
\end{gather}
\begin{gather}\label{lower-bounds}
\underline{M_{ijrs}} := 
\left\{
    \begin{array}{ll}
        0,  & \mbox{if } A_{ij} = 0 \\
        A_{ij} ( 1 - \log{A_{ij}} + \log{\tfrac{k_{i} k_{j}}{2 m}} ), & \mbox{if } A_{ij} \neq 0 \,.
    \end{array}
\right.
\end{gather}
For numerical stability, we also set a lower bound $\omega_{rs}^L = 10^{-12}$ since function $f_{ij}$ is not defined at $\omega_{rs} = 0$.

\subsection{Symmetry-breaking Constraints}

In the formulations discussed above, any permutation of the group indices in the community assignment variables $\mathbf{Z}$ leads to an equivalent solution.
Thus, each solution is, in practice, represented $K!$ times in the model, leading to an inefficient solution process.
To circumvent this issue, \citet{plastria2002formulating} proposed a set of linear constraints that limits the set of feasible solutions by eliminating solutions which are equivalent.
This is done by enforcing the model to accept only \textit{lexicographically minimal solutions}, i.e., by forcing community $r$ to always contain the lowest numbered object (vertex) which does not belong to any of the previous communities $1, \dots, r-1$.
As shown in~\cite{plastria2002formulating}, this can be achieved by including the following constraints:
\begin{equation}
    z_{11} = 1
\end{equation}
\begin{equation}
\sum_{i=2}^{j-1} \sum_{l=1}^{r-1} z_{il} - \sum_{l=1}^{r} z_{jl} \leq j - 3 \qquad \forall r \in \{2, \dots, K-1\}, \forall j \in \{r, \dots, n\}
\end{equation}
The last cluster $K$ is not associated with any constraint, as it will automatically contain all remaining objects which do not belong to any of the previous clusters.
These constraints effectively reduce the symmetry of the problem and speed-up the solution method.

\section{Heuristics for the DCSBM}
\label{section-heuristic}

The mathematical programming approaches discussed in Section \ref{section-exact} permit to find optimal solutions of the MLE model with a certificate of global optimality. 
In practice, however, they are limited to fairly small problems since their computational effort quickly rises as a function of the number of nodes in the graph. In contrast, heuristic methods usually solve a problem in reduced computational time but do not provide solution-quality guarantees.
We review three natural variants of the EM algorithm based on the method proposed by \citet{Karrer_2011} for community detection using the DCSBM.
These approaches are based on a local search heuristic on the space of community assignments. Since we can now find optimal solutions, we conduct a disciplined experimental analysis of these heuristics to measure how far they are from the known optima of the MLE model.

\subsection{Expectation-Maximization Algorithm}

The EM algorithm was introduced by \citet{dempster1977maximum} as a general iterative scheme for finding the parameter estimates of maximum likelihood or maximum a posteriori probability (MAP) of statistical models with (unobserved) latent variables.
The EM algorithm has been applied to a variety of models in machine learning, including Gaussian mixture models (GMM) and hidden Markov models (HMM)~\cite{xu1996convergence}, as well as data clustering~\cite{bottou1995convergence,jain2010data}. 
The algorithm iteratively performs an expectation step (E-step) and a maximization step (M-step).
In the context of community detection, the E-step searches for community assignments $\mathbf{Z}$ that maximize the likelihood function given a connectivity matrix $\boldsymbol{\Omega}$, whereas the M-step estimates $\boldsymbol{\Omega}$ using the current assignments $\mathbf{Z}$.
Each step is guaranteed to increase the log-likelihood function, and therefore the method converges towards a local optimum.
We describe the M-step in Section \ref{subsection-M-step} and discuss three natural algorithmic variations for the E-step in Section~\ref{subsection-E-step}.

\subsection{Maximization Step (M-step)}
\label{subsection-M-step}
The maximization step consists in estimating the parameters $\boldsymbol{\Omega}$ which maximize the likelihood function given a fixed assignment of vertices to communities $\mathbf{Z}$. 
The optimal value for $\omega_{rs}$ can be calculated in closed form as:
\begin{equation}\label{maximization-step}
\omega_{rs}^* = 2m \left(\frac{\sum_{i,j} A_{ij} z_{ir} z_{js}}{\sum_{i,j} k_i k_j z_{ir} z_{js}}\right) = \dfrac{2m \cdot m_{rs}}{\kappa_r \kappa_s} 
\end{equation}
where $m_{rs} = \sum_{i,j}^{n} A_{ij} z_{ir} z_{js}$ is the number of edges between groups $r$ and $s$, and $\kappa_r = \sum_{i}^{n} k_i z_{ir}$ is the sum of the degrees of the vertices in group $r$.

\subsection{Expectation Step (E-step)} 
\label{subsection-E-step}
The expectation step consists in searching for community assignments $\mathbf{Z}$ that maximize the likelihood given the current affinity matrix $\boldsymbol{\Omega}$.
This step corresponds to an $\mathcal{NP}$-hard combinatorial optimization problem~\cite{amini2013pseudo}.
There are different possible ways to perform the E-step. 
We highlight three main approaches which, combined with the M-step, result in three variations of the EM algorithm reported in Algorithms~\ref{em-ls1-algo}, \ref{em-ls2-algo}, and \ref{em-exact-algo}. In all cases, random community assignments are used as initial state.

\begin{multicols}{2}
\footnotesize{
\begin{algorithm}[H]
\setstretch{1.35}
\SetAlgoLined
\SetKwFunction{FuncMaximizationStep}{MaximizationStep}
Initialize community assignments $\mathbf{Z}$\;
\Repeat{\textnormal{The likelihood function can no longer be improved}}{
    $\boldsymbol{\Omega} \gets$ \FuncMaximizationStep{$\mathbf{Z}$}\;
    $\mathcal{L} \gets \log P(A|\boldsymbol{\Omega}, \mathbf{Z})$\;
    \Repeat{\textnormal{No improving relocation can be found}}{
        \For{\textnormal{each vertex $i \in V$ and group $r \in \mathcal{C}$}}{
            Consider $\mathbf{Z}'$ constructed from $\mathbf{Z}$ by relocating vertex $i$ to group $r$\;
            $\mathcal{L}' \gets \log P(A|\boldsymbol{\Omega}, \mathbf{Z}')$\; 
            \If{$\mathcal{L}' > \mathcal{L}$}{
                Apply relocation and update solution: \\ $\mathbf{Z} \gets \mathbf{Z}'$; $\mathcal{L} \gets \mathcal{L}'$\; 
            }
        }
    }
}
\caption{\texttt{EM-LS1} algorithm}
\label{em-ls1-algo}
\end{algorithm}
}

\columnbreak

\footnotesize{
\begin{algorithm}[H]
\setstretch{1.35}
\SetAlgoLined
Initialize community assignments $\mathbf{Z}$\;
    $\boldsymbol{\Omega} \gets$ \FuncMaximizationStep{$\mathbf{Z}$}\;
    $\mathcal{L} \gets \log P(A|\boldsymbol{\Omega}, \mathbf{Z})$\;
\Repeat{\textnormal{The likelihood function can no longer be improved}}{
    \Repeat{\textnormal{No improving relocation can be found}}{
        \For{\textnormal{each vertex $i \in V$ and group $r \in \mathcal{C}$}}{
            Consider $\mathbf{Z}'$ constructed from $\mathbf{Z}$ by relocating vertex $i$ to group $r$\;
            $\boldsymbol{\Omega}' \gets$ \FuncMaximizationStep{$\mathbf{Z}'$}\;
            $\mathcal{L}' \gets \log P(A|\boldsymbol{\Omega}', \mathbf{Z}')$\; 
            \If{$\mathcal{L}' > \mathcal{L}$}{
                Apply relocation and update solution: \\
                $\mathbf{Z} \gets \mathbf{Z}'$;
                $\mathcal{L} \gets \mathcal{L}'$;
                $\boldsymbol{\Omega} \gets \boldsymbol{\Omega}'$\;
            }
        }
    }
}
\caption{\texttt{EM-LS2} algorithm}
\label{em-ls2-algo}
\end{algorithm}
}
\end{multicols}

{\centering
\begin{minipage}{0.7\linewidth}
\footnotesize{
\begin{algorithm}[H]
\setstretch{1.35}
\SetAlgoLined
\SetKwFunction{FuncEexact}{E-exact}
Initialize community assignments $\mathbf{Z}$\;
\Repeat{\textnormal{The likelihood function can no longer be improved}}{
    $\boldsymbol{\Omega} \gets$ \FuncMaximizationStep{$\mathbf{Z}$}\;
    $\mathbf{Z} \gets $ \FuncEexact($\boldsymbol{\Omega}$) (Find $\mathbf{Z}$ by solving Model (\ref{expectation-exact-start})--(\ref{expectation-exact-end}))
}
\caption{\texttt{EM-exact} algorithm}
\label{em-exact-algo}
\end{algorithm}
}
\end{minipage}
\par
}

\paragraph{\texttt{EM-LS1} Algorithm: Local search on the community assignment variables}
The first variant, based on a local search approach, is described in Algorithm~\ref{em-ls1-algo}. 
Line 3 of the algorithm performs the M-step, while lines 5--12 correspond to the first variant of the E-step.
For a fixed value of $\boldsymbol{\Omega}$, the method searches for community assignments $\mathbf{Z}$ by iterating over each vertex $i$ of the graph and relocating it to a different community $r$ whenever it leads to an improvement in the likelihood function.
This procedure is repeated until no more improving relocation exists.
It is important to note that $\boldsymbol{\Omega}$ stays fixed during the improvement phase based on relocation (E-step) and is only optimized in the M-step.

\paragraph{\texttt{EM-LS2} Algorithm: Local search integrated with M-step}
This variant is described in Algorithm \ref{em-ls2-algo}.
Here the value of $\boldsymbol{\Omega}$ is re-estimated (with the M-step) each time a vertex relocation is evaluated.
Therefore, the maximization step is ``embedded" into the expectation step (line~8).
Calculating the value of the likelihood function can be done in $O(K^2 n^2)$ elementary operations from scratch.
Yet, \citet{Karrer_2011} described how to find the best relocation move more efficiently by instead evaluating the change in the likelihood, exploiting the property that, when a vertex changes groups, only some terms of the likelihood function need to be updated. Thus, finding the community relocation that produces the maximum increase in the likelihood function can be done in time $O(K (K + \bar{k}))$ on average, where $\bar{k}$ is the average degree of a vertex in $G$.

\paragraph{\texttt{EM-exact} Algorithm: Exact community assignments}
Finally, as shown in Algorithm~\ref{em-exact-algo}, the complete E-step can be formulated as an integer quadratic program (IQP) of Equations (\ref{expectation-exact-start}--\ref{expectation-exact-end}) and solved using an exact solution method. 
\begin{align}\label{expectation-exact-start}
& \underset{\mathbf{Z}}{\text{minimize}}
& & \dfrac{1}{2} \sum_{i,j}^{n} \sum_{r,s}^{K} f_{ij}(\omega_{rs}) \, z_{ir} z_{js} \\
& \text{subject to}
& & \sum_{r=1}^{q} z_{ir} = 1 && \forall i \in V \\
& & & z_{ir} \in \{0, 1\} && \forall i \in V, r \in \mathcal{C} \label{expectation-exact-end}
\end{align}
This variant of EM effectively applies, in turn, an optimal expectation step and an optimal maximization step. It is, therefore, the approach that is closest to the canonical EM concept. Model (\ref{expectation-exact-start})--(\ref{expectation-exact-end}) seeks community assignments $\mathbf{Z}$ that maximize the likelihood, for a fixed $\boldsymbol{\Omega}$. 
In contrast to Model (\ref{descriptive-formulation-start})--(\ref{descriptive-formulation-end}), the terms $f_{ij}(\omega_{rs})$ in the objective function are now constant. Despite this simplification, the E-step remains an $\mathcal{NP}$-hard problem~\cite{amini2013pseudo}. It can be solved to optimality using standard MIP solvers based on branch-and-cut, such as \texttt{Gurobi} and \texttt{CPLEX}, for small and medium instances. 
It is less scalable, but noteworthy as a benchmark to evaluate the impact of optimal expectation steps in EM heuristics.
This variant of the EM algorithm effectively becomes a \textit{matheuristic}~\cite{boschetti2009matheuristics}, a term used to refer to methods that combine  metaheuristics with mathematical programming components.

\section{Computational Experiments}
\label{section-experiments}

The goals of our computational experiments are twofold.
\begin{enumerate}
    \item We compare the performance of the proposed exact methods in terms of computational time and scalability.
    \item Using our knowledge of optimal solutions and bounds, we measure to which extent the heuristics can find the true optimum of the maximum likelihood estimation problem.
\end{enumerate}

The experiments were performed on an Intel Xeon E5-2620 2.1 GHz processor machine with 128~GB of RAM and CentOS Linux 7 (Core) operating system.
The high-level programming language used in the implementation was \texttt{Julia}~\cite{bezanson2017julia}, and the package \texttt{JuMP}~\cite{dunning2017jump} was used as the modeling language for the exact methods.
The underlying optimization solvers adopted for the exact methods were \texttt{Couenne}~\cite{belotti2009couenne} as the global optimization solver for the MINLP and \texttt{CPLEX} for the MILP. For reproducibility, we provide our source code and all experimental data at \url{github.com/vidalt/Optimal-SBM}.

\subsection{Instances}

Synthetic graphs allow us to control the factors that might influence the difficulty of community detection, such as network size and community structure (e.g., degree of separability and assortativity).
Therefore, we generated two groups of data sets, denoted S1 and S2, composed of synthetic graphs produced by the DCSBM.
These graphs contain a number of vertices $n$ ranging from 8 to 16 and a number of edges $m$ ranging from 4 to 115. This problem scale allows to find optimal solutions and, in the largest cases, still challenges the solution capabilities of the exact methods.

For group S1, we set $K=2$ and generated graphs with $n \in \{8, 10, 12, 14, 16\}$. Since $K=2$, the affinity matrix $\boldsymbol{\Omega}$ of the model has three parameters: two diagonal elements $\omega_{11}, \omega_{22}$ and one off-diagonal $\omega_{12} = \omega_{21}$.
For each $(\omega_{\text{in}}, \omega_{\text{out}}) \in \{0.1, 0.4, 0.6, 0.9\}^2$, such that $\omega_{\text{in}} \neq \omega_{\text{out}}$, we sampled $\omega_{11}, \omega_{22}$ from $\mathcal{U}(\omega_{\text{in}} - 0.1, \omega_{\text{in}} + 0.1)$ and we sampled $\omega_{12}$ from $\mathcal{U}(\omega_{\text{out}} - 0.1, \omega_{\text{out}} + 0.1)$, where $\mathcal{U}(a,b)$ is the uniform distribution in the interval $[a,b]$.
This gives $4 \times 3=12$ combinations of values for $(\omega_{\text{in}}, \omega_{\text{out}})$.
Six combinations are assortative (with $\omega_{\text{in}} > \omega_{\text{out}}$) and six combinations are disassortative (with $\omega_{\text{in}} < \omega_{\text{out}}$). 
For statistical significance, we generated 10 instances for each combination, yielding a total of $5\times4\times3\times10 = 600$ instances.

Data group S2 is composed of strongly assortative graphs with $K \in \{2, 3\}$, $n \in \{8, 10,  12,  14, 16\}$ and three levels of community strength: low, medium and high. For each level of community strength, we sampled the diagonal and off-diagonal elements of $\boldsymbol{\Omega}$ from a uniform distribution in the corresponding interval given by Table~\ref{exp2-community-strength-def}. 
For each configuration, we generated a total of 10 instances, leading to $2\times3\times5\times10=300$ instances.

\begin{table}[htbp]
\centering
\setlength{\tabcolsep}{8pt}
\renewcommand{\arraystretch}{1.5}
\small{
\begin{tabular}{lllll}
\cline{2-4}
\multicolumn{1}{l|}{}         & \multicolumn{1}{l|}{LOW}            & \multicolumn{1}{l|}{MEDIUM}         & \multicolumn{1}{l|}{HIGH}           &  \\ \cline{1-4}
\multicolumn{1}{|l|}{$\omega_{rr}$} & \multicolumn{1}{l|}{{[}0.4, 1.0{]}} & \multicolumn{1}{l|}{{[}0.6, 1.0{]}} & \multicolumn{1}{l|}{{[}0.8, 1.0{]}} &  \\ \cline{1-4}
\multicolumn{1}{|l|}{$\omega_{rs} \, (r \neq s)$}  & \multicolumn{1}{l|}{{[}0.2, 0.4{]}} & \multicolumn{1}{l|}{{[}0.1, 0.3{]}} & \multicolumn{1}{l|}{{[}0.0, 0.2{]}} & \\ \cline{1-4}
\end{tabular}
}
\caption{Range of possible values for the diagonal and off-diagonal elements of the affinity matrix $\boldsymbol{\Omega}$}
\label{exp2-community-strength-def}
\end{table}

\subsection{Performance of the exact methods}

For each instance in S1 and S2, we run the two exact methods (MINLP and MILP) with a time limit of 600 seconds. 
To assess the impact of the symmetry-breaking constraints (SBC) in the solution time, we run each method twice: with (SBC) and without them (NoSBC).\\

\noindent
\textbf{General Comparison.}
Tables \ref{exp1-exact_results} and \ref{exp2-exact_results} present the following results for the exact methods: number of instances solved to optimality (``Opt''), percentage gap (``Gap''), solution time in seconds (``Time"), and number of nodes explored in the branch-and-bound tree (``Nodes'').
The values reported in both tables are averaged over the 10 instances of each type (except for ``Opt'').
The exact methods' percentage gap is calculated based on the log-likelihood function (including the constant terms) as ${\text{Gap} = (\text{UB} - \text{LB})/\text{UB}}$, where LB and UB are the lower and upper objective bounds.
A summary line is included in the bottom of each table showing the total number of instances solved to optimality (for column ``Opt") and average results for all other columns.

\begin{table}[htbp]
\centering
\setlength{\tabcolsep}{4pt}
\vspace*{-1cm}
\hspace*{-1.3cm}
{\fontsize{9}{10.5}\selectfont
\begin{tabular}{|c|cc|rrrr|rrrr|rrrr|rrrr|}
\hline
   &  &  & \multicolumn{8}{c|}{\textbf{MINLP}} & \multicolumn{8}{c|}{\textbf{MILP}} \\
   & &  & \multicolumn{4}{c|}{NoSBC} & \multicolumn{4}{c|}{SBC} & \multicolumn{4}{c|}{NoSBC} & \multicolumn{4}{c|}{SBC} \\
   $n$ & $\omega_{\text{in}}$ & $\omega_{\text{out}}$ &   Opt & Gap & Time & Nodes & Opt & Gap & Time & Nodes & Opt & Gap & Time & Nodes &   Opt & Gap & Time & Nodes \\
\hline
8  & 0.1 & 0.4 &    10 &    0.00 &      6.3 &     306.3 &  10 &    0.00 &      3.7 &     167.6 &    10 &    0.00 &      0.8 &      92.5 &  10 &    0.00 &      0.6 &      59.6 \\
   &     & 0.6 &    10 &    0.00 &     10.1 &     587.4 &  10 &    0.00 &      6.6 &     274.7 &    10 &    0.00 &      1.2 &     166.6 &  10 &    0.00 &      0.8 &      86.9 \\
   &     & 0.9 &    10 &    0.00 &     11.0 &     569.3 &  10 &    0.00 &      7.0 &     216.7 &    10 &    0.00 &      1.2 &     120.0 &  10 &    0.00 &      0.8 &      60.3 \\
   & 0.4 & 0.1 &    10 &    0.00 &      6.5 &     388.5 &  10 &    0.00 &      4.2 &     203.8 &    10 &    0.00 &      0.9 &     140.7 &  10 &    0.00 &      0.7 &      69.4 \\
   &     & 0.6 &    10 &    0.00 &     12.9 &     730.0 &  10 &    0.00 &      7.8 &     349.5 &    10 &    0.00 &      2.0 &     298.5 &  10 &    0.00 &      1.2 &     146.4 \\
   &     & 0.9 &    10 &    0.00 &     14.9 &     661.3 &  10 &    0.00 &      9.7 &     317.6 &    10 &    0.00 &      2.1 &     271.0 &  10 &    0.00 &      1.3 &     138.4 \\
   & 0.6 & 0.1 &    10 &    0.00 &      8.2 &     459.6 &  10 &    0.00 &      6.0 &     266.7 &    10 &    0.00 &      1.1 &     157.4 &  10 &    0.00 &      0.8 &     104.5 \\
   &     & 0.4 &    10 &    0.00 &     14.4 &     704.2 &  10 &    0.00 &      9.2 &     306.8 &    10 &    0.00 &      2.0 &     302.5 &  10 &    0.00 &      1.2 &     148.2 \\
   &     & 0.9 &    10 &    0.00 &     19.5 &     796.1 &  10 &    0.00 &     11.9 &     326.3 &    10 &    0.00 &      2.8 &     340.7 &  10 &    0.00 &      1.7 &     171.6 \\
   & 0.9 & 0.1 &    10 &    0.00 &     10.2 &     463.8 &  10 &    0.00 &      6.9 &     238.5 &    10 &    0.00 &      1.0 &     126.6 &  10 &    0.00 &      0.8 &      73.2 \\
   &     & 0.4 &    10 &    0.00 &     16.1 &     786.1 &  10 &    0.00 &     10.4 &     333.5 &    10 &    0.00 &      2.4 &     359.6 &  10 &    0.00 &      1.5 &     179.0 \\
   &     & 0.6 &    10 &    0.00 &     16.5 &     797.9 &  10 &    0.00 &     10.8 &     381.1 &    10 &    0.00 &      2.6 &     359.8 &  10 &    0.00 &      1.6 &     210.5 \\
\hline
10 & 0.1 & 0.4 &    10 &    0.00 &     27.4 &    2056.3 &  10 &    0.00 &     15.4 &     915.9 &    10 &    0.00 &      3.7 &     608.7 &  10 &    0.00 &      2.6 &     311.7 \\
   &     & 0.6 &    10 &    0.00 &     28.6 &    1676.5 &  10 &    0.00 &     17.7 &     839.2 &    10 &    0.00 &      3.6 &     440.6 &  10 &    0.00 &      2.3 &     223.7 \\
   &     & 0.9 &    10 &    0.00 &     39.4 &    2143.0 &  10 &    0.00 &     22.6 &     999.1 &    10 &    0.00 &      4.2 &     397.1 &  10 &    0.00 &      2.6 &     192.2 \\
   & 0.4 & 0.1 &    10 &    0.00 &     22.7 &    1585.0 &  10 &    0.00 &     15.9 &     988.8 &    10 &    0.00 &      3.1 &     514.9 &  10 &    0.00 &      2.1 &     301.7 \\
   &     & 0.6 &    10 &    0.00 &     49.5 &    3625.7 &  10 &    0.00 &     28.8 &    1528.7 &    10 &    0.00 &      8.5 &    1347.8 &  10 &    0.00 &      5.4 &     672.1 \\
   &     & 0.9 &     9 &    0.00 &    106.9 &   17598.8 &  10 &    0.00 &     31.9 &    1433.4 &    10 &    0.00 &      9.0 &    1198.6 &  10 &    0.00 &      5.9 &     584.7 \\
   & 0.6 & 0.1 &    10 &    0.00 &     35.8 &    2427.8 &  10 &    0.00 &     21.8 &    1140.8 &    10 &    0.00 &      4.1 &     720.2 &  10 &    0.00 &      2.8 &     370.5 \\
   &     & 0.4 &     9 &    0.00 &     99.8 &   18520.7 &  10 &    0.00 &     27.2 &    1243.0 &    10 &    0.00 &      7.1 &     976.3 &  10 &    0.00 &      4.9 &     568.6 \\
   &     & 0.9 &    10 &    0.00 &     57.9 &    3393.3 &  10 &    0.00 &     35.8 &    1563.7 &    10 &    0.00 &     12.8 &    1561.5 &  10 &    0.00 &      7.9 &     783.8 \\
   & 0.9 & 0.1 &    10 &    0.00 &     41.4 &    2126.2 &  10 &    0.00 &     23.7 &     942.5 &    10 &    0.00 &      4.6 &     545.2 &  10 &    0.00 &      3.0 &     300.1 \\
   &     & 0.4 &    10 &    0.00 &     57.2 &    3238.4 &  10 &    0.00 &     32.3 &    1427.1 &    10 &    0.00 &      8.4 &    1164.3 &  10 &    0.00 &      5.5 &     562.8 \\
   &     & 0.6 &    10 &    0.00 &     59.1 &    3178.2 &  10 &    0.00 &     33.5 &    1528.7 &    10 &    0.00 &     12.1 &    1610.8 &  10 &    0.00 &      7.1 &     707.6 \\
\hline
12 & 0.1 & 0.4 &    10 &    0.00 &     86.0 &    6522.3 &  10 &    0.00 &     48.8 &    3280.1 &    10 &    0.00 &     11.3 &    1955.8 &  10 &    0.00 &      7.0 &     946.8 \\
   &     & 0.6 &    10 &    0.00 &    105.6 &    7322.4 &  10 &    0.00 &     59.9 &    3242.9 &    10 &    0.00 &     11.1 &    1523.8 &  10 &    0.00 &      7.7 &     769.8 \\
   &     & 0.9 &    10 &    0.00 &    130.5 &    8374.4 &  10 &    0.00 &     70.3 &    3716.4 &    10 &    0.00 &     13.6 &    1551.8 &  10 &    0.00 &     10.1 &    1042.3 \\
   & 0.4 & 0.1 &    10 &    0.00 &     98.2 &    8169.0 &  10 &    0.00 &     51.0 &    3613.0 &    10 &    0.00 &     10.4 &    2138.1 &  10 &    0.00 &      7.0 &    1222.5 \\
   &     & 0.6 &    10 &    0.00 &    194.6 &   14369.8 &  10 &    0.00 &    102.1 &    6625.3 &    10 &    0.00 &     37.8 &    4945.8 &  10 &    0.00 &     23.0 &    2786.3 \\
   &     & 0.9 &    10 &    0.00 &    176.7 &   11083.3 &  10 &    0.00 &     99.9 &    5439.4 &    10 &    0.00 &     36.0 &    4215.0 &  10 &    0.00 &     21.6 &    2205.0 \\
   & 0.6 & 0.1 &    10 &    0.00 &    120.5 &    9293.9 &  10 &    0.00 &     65.4 &    4137.3 &    10 &    0.00 &     14.5 &    2603.0 &  10 &    0.00 &      8.1 &    1166.3 \\
   &     & 0.4 &    10 &    0.00 &    169.5 &   13084.0 &  10 &    0.00 &     96.6 &    6478.6 &    10 &    0.00 &     35.0 &    5318.6 &  10 &    0.00 &     19.4 &    2735.3 \\
   &     & 0.9 &    10 &    0.00 &    206.1 &   14039.0 &  10 &    0.00 &    109.4 &    6667.8 &    10 &    0.00 &     49.4 &    5938.7 &  10 &    0.00 &     27.8 &    3003.1 \\
   & 0.9 & 0.1 &    10 &    0.00 &    148.8 &    9853.2 &  10 &    0.00 &     78.9 &    4347.5 &    10 &    0.00 &     21.0 &    2808.6 &  10 &    0.00 &     12.1 &    1266.2 \\
   &     & 0.4 &    10 &    0.00 &    177.3 &   11626.7 &  10 &    0.00 &     96.7 &    5881.0 &    10 &    0.00 &     33.5 &    4288.4 &  10 &    0.00 &     18.9 &    2170.4 \\
   &     & 0.6 &     9 &    0.00 &    258.6 &   24597.3 &  10 &    0.00 &    111.4 &    6751.7 &    10 &    0.00 &     46.4 &    5334.6 &  10 &    0.00 &     26.9 &    2711.4 \\
\hline
14 & 0.1 & 0.4 &     9 &    0.00 &    362.9 &   30388.0 &  10 &    0.00 &    188.2 &   14498.9 &    10 &    0.00 &     35.1 &    5687.8 &  10 &    0.00 &     22.1 &    3263.6 \\
   &     & 0.6 &    10 &    0.00 &    445.0 &   33400.6 &  10 &    0.00 &    230.4 &   16568.3 &    10 &    0.00 &     46.8 &    7168.1 &  10 &    0.00 &     23.8 &    3236.5 \\
   &     & 0.9 &     9 &    0.40 &    420.9 &   27246.6 &  10 &    0.00 &    215.2 &   12588.9 &    10 &    0.00 &     45.3 &    4864.8 &  10 &    0.00 &     29.2 &    2746.2 \\
   & 0.4 & 0.1 &     8 &    0.49 &    478.5 &   37337.6 &  10 &    0.00 &    233.9 &   17723.4 &    10 &    0.00 &     58.2 &   10053.6 &  10 &    0.00 &     28.1 &    4597.2 \\
   &     & 0.6 &     1 &    4.85 &    574.6 &   37600.6 &  10 &    0.00 &    350.0 &   24600.9 &    10 &    0.00 &    180.8 &   19339.3 &  10 &    0.00 &     98.0 &   10378.5 \\
   &     & 0.9 &     1 &    6.86 &    594.4 &   35976.6 &  10 &    0.00 &    395.0 &   24594.2 &    10 &    0.00 &    217.5 &   18671.0 &  10 &    0.00 &    111.4 &    9342.6 \\
   & 0.6 & 0.1 &     8 &    0.63 &    425.6 &   30829.4 &  10 &    0.00 &    235.4 &   16307.2 &    10 &    0.00 &     52.7 &    7601.7 &  10 &    0.00 &     29.0 &    3664.9 \\
   &     & 0.4 &     2 &    5.95 &    598.5 &   39127.7 &  10 &    0.00 &    353.0 &   25993.2 &    10 &    0.00 &    152.0 &   16275.0 &  10 &    0.00 &     75.2 &    7830.8 \\
   &     & 0.9 &     1 &    9.92 &    596.3 &   33486.5 &  10 &    0.00 &    442.1 &   28049.6 &    10 &    0.00 &    252.1 &   21144.3 &  10 &    0.00 &    130.3 &    9842.1 \\
   & 0.9 & 0.1 &     8 &    0.51 &    530.5 &   35931.5 &  10 &    0.00 &    254.7 &   15866.8 &    10 &    0.00 &     62.8 &    6718.1 &  10 &    0.00 &     33.2 &    3399.7 \\
   &     & 0.4 &     1 &    7.86 &    598.8 &   36665.6 &  10 &    0.00 &    398.0 &   26503.8 &    10 &    0.00 &    200.2 &   18398.8 &  10 &    0.00 &    115.6 &   11125.2 \\
   &     & 0.6 &     1 &   11.35 &    597.1 &   32256.1 &  10 &    0.00 &    426.7 &   25344.6 &    10 &    0.00 &    236.0 &   17489.5 &  10 &    0.00 &    122.0 &    9803.4 \\
\hline
16 & 0.1 & 0.4 &     0 &   15.88 &    601.0 &   33644.3 &   1 &    9.23 &    592.4 &   38779.1 &    10 &    0.00 &    262.8 &   35403.1 &  10 &    0.00 &    122.7 &   17711.9 \\
   &     & 0.6 &     0 &   19.67 &    600.9 &   31410.4 &   2 &    8.27 &    573.8 &   32368.4 &    10 &    0.00 &    141.1 &   16431.6 &  10 &    0.00 &     88.5 &    9795.2 \\
   &     & 0.9 &     0 &   23.84 &    600.8 &   26570.6 &   2 &   11.29 &    575.4 &   27781.8 &    10 &    0.00 &    165.7 &   12670.4 &  10 &    0.00 &     93.0 &    6962.4 \\
   & 0.4 & 0.1 &     1 &   13.86 &    575.5 &   34141.5 &   1 &    7.31 &    557.7 &   36824.9 &     9 &    0.58 &    247.0 &   31551.6 &  10 &    0.00 &    140.8 &   17516.1 \\
   &     & 0.6 &     0 &   29.13 &    600.8 &   26345.5 &   0 &   18.02 &    600.8 &   29144.8 &     1 &    6.30 &    579.9 &   39058.6 &   9 &    0.32 &    471.0 &   33429.2 \\
   &     & 0.9 &     0 &   31.66 &    600.6 &   22139.2 &   0 &   21.77 &    600.6 &   24664.4 &     3 &    4.97 &    528.6 &   31429.7 &   8 &    0.87 &    416.9 &   25972.9 \\
   & 0.6 & 0.1 &     0 &   15.10 &    600.8 &   31180.2 &   4 &    7.95 &    562.1 &   33916.7 &     8 &    0.81 &    282.0 &   30419.5 &   9 &    0.28 &    170.1 &   18015.5 \\
   &     & 0.4 &     0 &   23.83 &    600.8 &   25580.5 &   0 &   16.84 &    600.9 &   30136.4 &     1 &    5.16 &    599.7 &   47506.8 &   9 &    0.28 &    417.5 &   35457.5 \\
   &     & 0.9 &     0 &   31.91 &    600.6 &   19998.5 &   0 &   24.05 &    600.8 &   24056.4 &     0 &    7.70 &    600.0 &   32472.5 &   5 &    2.56 &    561.4 &   33329.6 \\
   & 0.9 & 0.1 &     0 &   19.38 &    600.7 &   27602.0 &   3 &    8.96 &    558.4 &   28266.1 &     9 &    0.47 &    177.3 &   16306.2 &  10 &    0.00 &    109.6 &    9363.8 \\
   &     & 0.4 &     0 &   27.95 &    600.6 &   21412.2 &   0 &   21.98 &    600.7 &   24294.1 &     1 &    6.59 &    592.3 &   35953.7 &   8 &    0.78 &    484.6 &   32075.5 \\
   &     & 0.6 &     0 &   30.05 &    600.5 &   21792.6 &   0 &   21.64 &    600.6 &   25592.2 &     0 &    9.43 &    600.0 &   30829.3 &   4 &    3.29 &    585.3 &   32882.9 \\
\hline
\multicolumn{3}{|c|}{Aggregate} & 417 & 5.52 & 267.5 & 15653.7 & 493 & 2.96 & 202.3 & 11376.3 & 542 & 0.70 & 112.3 & 9498.2 & 582 & 0.14 & 78.9 & 6346.6 \\
\hline
\end{tabular}
}
\caption{General performance of the exact methods (MINLP vs MILP) for data group S1, with SBC and without them (NoSBC)}
\label{exp1-exact_results}
\end{table}

\begin{table}[htbp]
\centering
\setlength{\tabcolsep}{4pt}
\hspace*{-1.7cm}
{\fontsize{8.6}{10.5}\selectfont
\begin{tabular}{|c|c|l|rrrr|rrrr|rrrr|rrrr|}
\hline
  & & & \multicolumn{8}{c|}{\textbf{MINLP}} & \multicolumn{8}{c|}{\textbf{MILP}} \\
  & & Comm. & \multicolumn{4}{c|}{NoSBC} & \multicolumn{4}{c|}{SBC} & \multicolumn{4}{c|}{NoSBC} & \multicolumn{4}{c|}{SBC} \\
  $K$ & $n$ & strength &   Opt & Gap & Time & Nodes & Opt & Gap & Time & Nodes & Opt & Gap & Time & Nodes & Opt & Gap & Time & Nodes \\
\hline
2 & 8  & low &    10 &    0.00 &   12.9 &    551.1 &  10 &   0.00 &    8.3 &    259.9 &    10 &   0.00 &    3.4 &    250.1 &  10 &   0.00 &    1.9 &    122.7 \\
  &    & medium &    10 &    0.00 &   12.8 &    726.3 &  10 &   0.00 &    8.6 &    309.3 &    10 &   0.00 &    3.3 &    268.8 &  10 &   0.00 &    1.9 &    131.8 \\
  &    & high &    10 &    0.00 &   10.1 &    432.4 &  10 &   0.00 &    6.4 &    204.2 &    10 &   0.00 &    2.2 &    142.7 &  10 &   0.00 &    1.4 &     72.1 \\
  & 10 & low &    10 &    0.00 &   42.5 &   2543.1 &  10 &   0.00 &   26.1 &   1147.1 &    10 &   0.00 &    5.9 &    835.6 &  10 &   0.00 &    3.9 &    428.7 \\
  &    & medium &    10 &    0.00 &   41.5 &   2717.2 &  10 &   0.00 &   25.1 &   1205.4 &    10 &   0.00 &    6.0 &    878.3 &  10 &   0.00 &    4.1 &    454.4 \\
  &    & high &    10 &    0.00 &   39.4 &   2260.7 &  10 &   0.00 &   24.7 &    985.2 &    10 &   0.00 &    4.3 &    529.6 &  10 &   0.00 &    2.7 &    270.7 \\
  & 12 & low &    10 &    0.00 &  163.0 &  11381.0 &  10 &   0.00 &   91.9 &   5779.5 &    10 &   0.00 &   49.9 &   5554.3 &  10 &   0.00 &   20.1 &   2414.6 \\
  &    & medium &    10 &    0.00 &  149.8 &  10292.3 &  10 &   0.00 &   81.5 &   4846.0 &    10 &   0.00 &   28.8 &   3401.1 &  10 &   0.00 &   11.9 &   1395.9 \\
  &    & high &    10 &    0.00 &  144.9 &   7602.1 &  10 &   0.00 &   73.1 &   3672.0 &    10 &   0.00 &   15.2 &   1924.3 &  10 &   0.00 &    6.9 &    817.4 \\
  & 14 & low &     4 &    3.92 &  575.2 &  38428.5 &  10 &   0.00 &  346.0 &  24528.2 &    10 &   0.00 &  153.3 &  12810.8 &  10 &   0.00 &   82.8 &   6691.3 \\
  &    & medium &     4 &    4.66 &  511.1 &  32873.6 &  10 &   0.00 &  301.8 &  21439.0 &    10 &   0.00 &  162.3 &  12307.9 &  10 &   0.00 &   88.1 &   7492.4 \\
  &    & high &    10 &    0.00 &  457.3 &  30102.3 &  10 &   0.00 &  246.4 &  15366.9 &    10 &   0.00 &   84.6 &   6258.7 &  10 &   0.00 &   38.3 &   3014.3 \\
  & 16 & low &     0 &   24.59 &  600.8 &  26048.4 &   0 &  18.47 &  600.9 &  30157.5 &     2 &   6.37 &  579.9 &  27377.8 &   6 &   1.63 &  474.1 &  25546.0 \\
  &    & medium &     0 &   23.94 &  600.8 &  25228.8 &   0 &  16.80 &  600.8 &  28751.8 &     3 &   5.57 &  544.9 &  17520.7 &   5 &   2.24 &  437.8 &  15999.2 \\
  &    & high &     0 &   21.02 &  600.7 &  27075.7 &   1 &   9.41 &  598.9 &  33362.2 &     7 &   2.24 &  376.9 &  12027.5 &  10 &   0.00 &  252.3 &   9534.4 \\
\hline
3 & 8  & low &     6 &    9.86 &  486.0 &   8260.7 &  10 &   0.00 &   65.8 &   1887.0 &    10 &   0.00 &   31.1 &   2505.8 &  10 &   0.00 &    9.3 &    520.3 \\
  &    & medium &     8 &    2.89 &  436.3 &  10404.9 &  10 &   0.00 &   51.0 &   1668.7 &    10 &   0.00 &   39.3 &   2866.0 &  10 &   0.00 &    8.0 &    410.2 \\
  &    & high &     6 &   10.93 &  404.7 &   6771.4 &  10 &   0.00 &   52.3 &   1788.2 &    10 &   0.00 &   22.8 &   1827.0 &  10 &   0.00 &    6.0 &    372.6 \\
  & 10 & low &     0 &   33.24 &  600.9 &  23751.9 &   9 &   0.00 &  361.3 &  20935.7 &    10 &   0.00 &  225.5 &  22935.7 &  10 &   0.00 &   37.1 &   4417.9 \\
  &    & medium &     3 &   18.32 &  532.7 &  25077.7 &   9 &   0.00 &  284.0 &  18346.7 &    10 &   0.00 &  106.6 &  11739.5 &  10 &   0.00 &   23.1 &   3193.6 \\
  &    & high &     3 &   15.26 &  490.3 &  28013.6 &  10 &   0.00 &  207.3 &  12059.5 &    10 &   0.00 &   67.3 &   6818.6 &  10 &   0.00 &   12.7 &   1653.1 \\
  & 12 & low &     0 &   60.62 &  601.0 &  16741.0 &   0 &  39.07 &  600.7 &  18220.6 &     0 &  10.17 &  600.0 &  32058.7 &  10 &   0.00 &  348.6 &  27153.8 \\
  &    & medium &     0 &   58.57 &  601.0 &  15707.1 &   0 &  33.35 &  600.6 &  17730.2 &     2 &   8.06 &  538.6 &  27221.1 &   9 &   0.36 &  299.5 &  20088.8 \\
  &    & high &     0 &   59.16 &  601.1 &  17991.1 &   0 &  31.70 &  600.6 &  20065.9 &     6 &   2.15 &  474.5 &  28412.3 &  10 &   0.00 &  115.5 &   9244.4 \\
  & 14 & low &     0 &   74.51 &  600.5 &   7937.1 &   0 &  69.01 &  600.2 &   8061.0 &     0 &  19.77 &  600.0 &  13066.7 &   0 &  13.02 &  600.0 &  16901.7 \\
  &    & medium &     0 &   64.45 &  600.5 &   7016.3 &   0 &  68.67 &  600.2 &   7958.9 &     0 &  18.89 &  600.0 &  12710.3 &   0 &  12.62 &  600.0 &  15150.3 \\
  &    & high &     0 &   57.07 &  600.7 &   9456.2 &   0 &  64.72 &  600.3 &   9602.9 &     0 &  11.25 &  600.0 &  17031.9 &   7 &   2.36 &  492.6 &  23093.9 \\
  & 16 & low &     0 &  100.00 &  599.8 &   1770.4 &   0 &  92.69 &  599.7 &   3017.5 &     0 &  29.09 &  600.0 &   4430.8 &   0 &  24.13 &  600.0 &   6488.1 \\
  &    & medium &     0 &  100.00 &  600.0 &   2407.4 &   0 &  92.67 &  599.6 &   2509.6 &     0 &  27.60 &  600.0 &   4882.4 &   0 &  22.48 &  600.0 &   6196.3 \\
  &    & high &     0 &   99.26 &  600.0 &   2587.7 &   0 &  93.27 &  599.9 &   4183.8 &     0 &  25.58 &  600.0 &   7191.3 &   0 &  19.77 &  600.0 &   7161.1 \\
\hline
\multicolumn{3}{|c|}{Aggregate} & 134 & 28.08 & 410.6 & 13405.3 & 179 & 20.99 & 315.5 & 10668.3 & 200 & 5.6 & 257.6 & 9926.2 & 237 & 3.29 & 192.7 & 7214.4 \\
\hline
\end{tabular}
}
\caption{General performance of the exact methods (MILP vs MINLP) for data group S2, with SBC and without them (NoSBC)}
\label{exp2-exact_results}
\end{table}

As seen in these results, the MILP is clearly faster than the MINLP for both groups of data sets. When $K=2$, the MILP can optimally solve all instances with up to $n=14$ vertices, whereas the MINLP is already unable to find the optimum for some instances with 14 vertices. 
This visible reduction in solution time is a consequence of a more efficient branch-and-bound exploration.
Indeed, for the instances that are solved to optimality, the MILP visits fewer search nodes than the MINLP.
For larger instances (e.g., when $K=3$ in Table~\ref{exp2-exact_results}), both methods have difficulties to consistently find optimal solutions within the time limit. However, for instances that could not be solved to optimality, percentage gaps are generally much smaller for the MILP. These results clearly illustrate that the DCSBM is indeed challenging to solve to optimality.\\

\begin{figure}[htbp]
\includegraphics[width=0.9\textwidth]{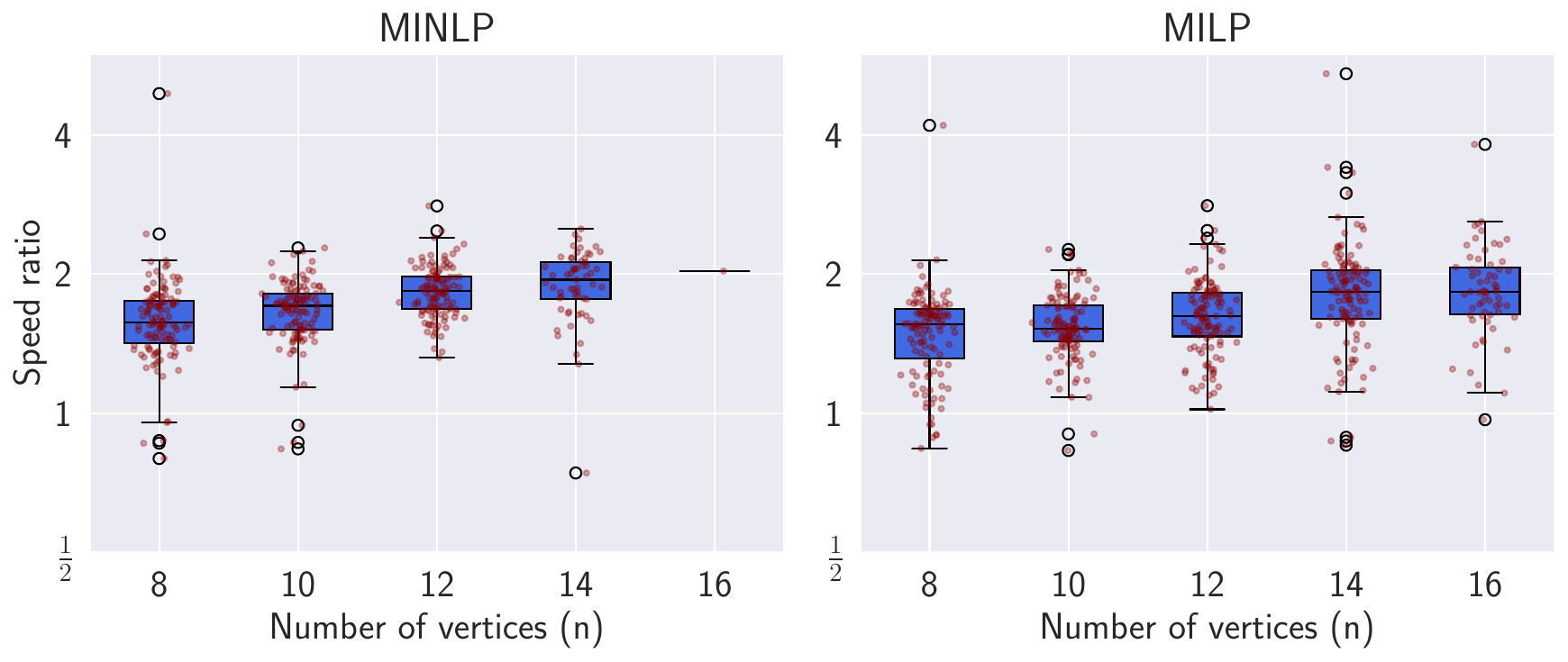}
\centering
\caption{Speed ratios between between the solution time of the methods \textit{without} and \textit{with} SBC (MINLP on the left and MILP on the right)}
\label{fig:symm_break_speed_ratio}
\end{figure}

\noindent
\textbf{Impact of the symmetry-breaking constraints.}
Figure~\ref{fig:symm_break_speed_ratio} illustrates the impact of using SBC with the MINLP (on the left) and with the MILP (on the right).
It shows the speed ratio between the solution time of the method without and with SBC, depending on the number of vertices in data group~S1.
The results are summarized as boxplots, with whiskers that extend to 1.5 times the interquartile range.
Points outside this range are marked as outliers and noted with a ``$\circ$".

As highlighted on Figure~\ref{fig:symm_break_speed_ratio}, the use of SBC has a beneficial impact on the MINLP and MILP solution methods. Even in the simple case with $K=2$, adding SBC clearly improves the solution time of both exact methods for the great majority of instances. The improvement becomes more marked as $n$ increases, with solution times up to 2x faster.\\

\begin{figure}[htbp]
\includegraphics[width=0.9\textwidth]{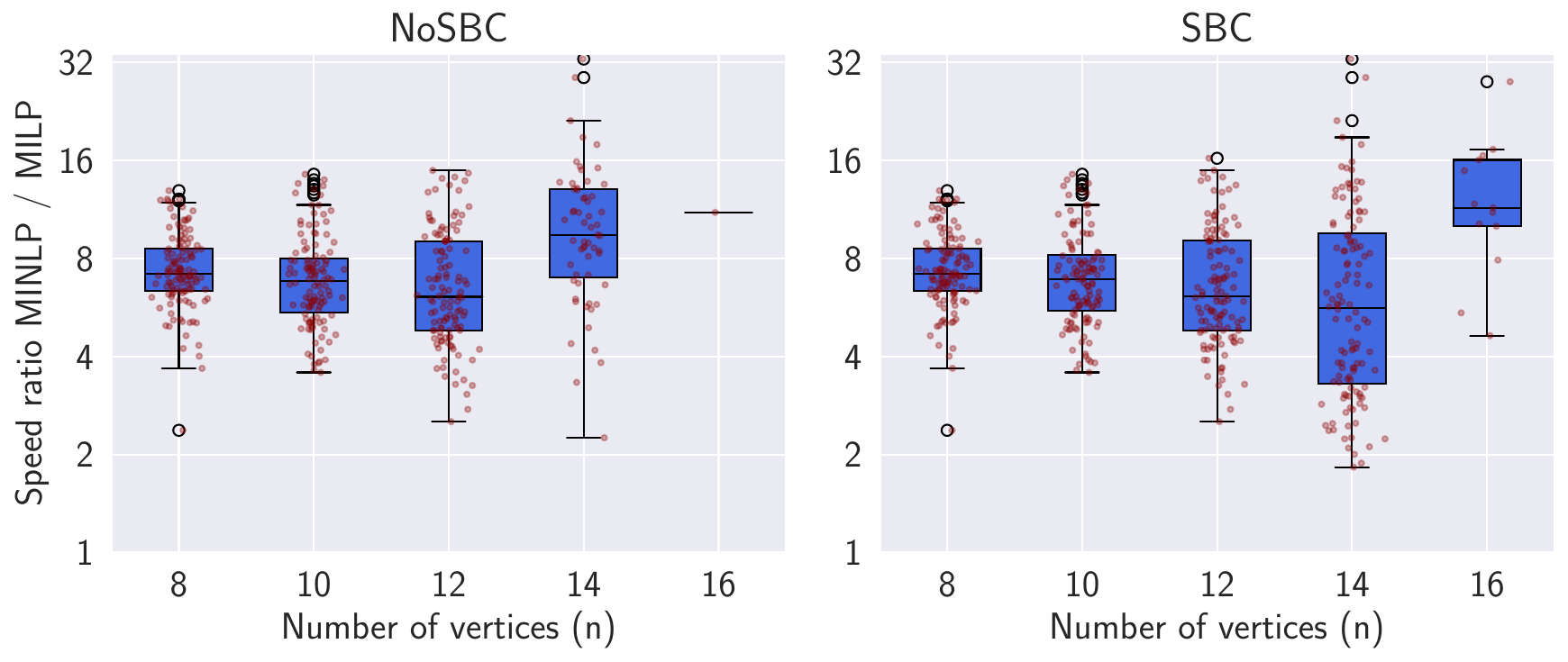}
\centering
\caption{Speed ratios between the solution time of the MINLP and MILP approaches (without SBC on the left, and with SBC on the right)}
\label{fig:speed_ratio}
\end{figure}

\noindent
\textbf{Comparison of the formulations.}
Finally, Figure~\ref{fig:speed_ratio} compares the solution times of the exact methods on data sets of group S1.
As $n$ increases, the variance in the distribution of the speed ratio increases. The speed ratios are nonetheless always greater than 1, meaning that the MILP approach is faster than the MINLP, regardless of the use of the SBC. In some cases, the MILP is as high as 32 times faster than the MINLP.

\subsection{Performance of the heuristic methods}

To evaluate the performance of the heuristic solution approaches, we compare their solutions to the solutions found by the exact methods.
For each instance in S1 and S2, we run the three EM variants for 50 trials (with different random starts).
For each instance the relative percentage gap is calculated as: 
\begin{equation}
    \text{Gap(\%)} = \frac{\text{OBJ} - \text{BKS}}{\text{BKS}}
\end{equation}
where OBJ is the objective value of the heuristic solution, and BKS (best-known solution) is the objective value of the optimal or best integer-feasible solution found by the MILP with SBC. 

Tables \ref{exp1-heuristics-avg-gap} and \ref{exp2-heuristics-avg-gap} present average gap values and solution times out of the 50 trials for data groups S1 and S2, respectively. As visible from these results, the solution times of \texttt{EM-LS1} and \texttt{EM-LS2} are generally close, with \texttt{EM-LS1} being slightly faster on average. The solution time of \texttt{EM-exact} is orders of magnitude higher than that of the other two heuristics since it involves the exact solution of a MIP during each expectation step.

\begin{table}[htbp]
\centering
\vspace*{-1cm}
{\fontsize{9}{10.5}\selectfont
\begin{tabular}{|c|cc|rr|rr|rr|}
\hline
& &  & \multicolumn{2}{c|}{\texttt{E-LS1}} & \multicolumn{2}{c|}{\texttt{E-LS2}} & \multicolumn{2}{c|}{\texttt{E-exact}} \\
$n$ & $\omega_{\text{in}}$ & $\omega_{\text{out}}$ & Gap (\%) &     Time (s) &         Gap (\%) &     Time (s) &          Gap (\%) &     Time (s) \\
\hline
8  & 0.1 & 0.4 &  3.13 &  \num{0.00016438817977905274} &  1.65 &  \num{0.00024672508239746086} &    1.76 &  \num{0.39295141506195064} \\
   &     & 0.6 &  4.29 &  \num{0.00012672948837280274} &  2.08 &   \num{0.0002034840583801269} &    3.09 &   \num{0.4792081890106202} \\
   &     & 0.9 &  7.18 &   \num{0.0001445293426513672} &  3.44 &  \num{0.00024064302444458006} &    4.33 &   \num{0.4296951575279236} \\
   & 0.4 & 0.1 &  3.94 &  \num{0.00012197065353393555} &  2.51 &  \num{0.00018711900711059572} &    3.29 &  \num{0.39924089384078976} \\
   &     & 0.6 &  1.97 &   \num{0.0001478843688964844} &  1.27 &  \num{0.00020576858520507816} &    1.65 &   \num{0.5387380781173705} \\
   &     & 0.9 &  3.35 &  \num{0.00015810108184814454} &  1.75 &     \num{0.00025140380859375} &    2.89 &   \num{0.5436676049232483} \\
   & 0.6 & 0.1 &  3.56 &  \num{0.00011538600921630861} &  2.26 &  \num{0.00019304609298706054} &    2.70 &   \num{0.4917164025306701} \\
   &     & 0.4 &  2.10 &  \num{0.00014452266693115235} &  1.39 &    \num{0.000214146614074707} &    1.83 &   \num{0.6113001971244811} \\
   &     & 0.9 &  2.87 &  \num{0.00014308214187622067} &  1.90 &  \num{0.00024411869049072266} &    2.12 &   \num{0.6274569787979126} \\
   & 0.9 & 0.1 &  6.13 &  \num{0.00013185548782348634} &  3.68 &  \num{0.00021552991867065428} &    4.46 &   \num{0.4593194913864136} \\
   &     & 0.4 &  1.96 &  \num{0.00014389801025390625} &  1.42 &   \num{0.0002247629165649414} &    1.56 &   \num{0.6037003073692322} \\
   &     & 0.6 &  2.00 &  \num{0.00014445877075195315} &  1.39 &   \num{0.0002203221321105957} &    1.67 &   \num{0.6248697929382324} \\
\hline
10 & 0.1 & 0.4 &  2.26 &   \num{0.0004623627662658692} &  1.41 &  \num{0.00037025308609008793} &    1.45 &   \num{0.7310159134864808} \\
   &     & 0.6 &  3.57 &  \num{0.00024674510955810546} &  1.76 &   \num{0.0003897652626037598} &    2.44 &   \num{0.7053591847419739} \\
   &     & 0.9 &  6.06 &  \num{0.00024270820617675786} &  4.13 &  \num{0.00039296627044677735} &    3.98 &   \num{0.6657761030197143} \\
   & 0.4 & 0.1 &  2.62 &   \num{0.0002299079895019531} &  2.15 &    \num{0.000341644287109375} &    2.11 &   \num{0.6671136450767517} \\
   &     & 0.6 &  1.37 &  \num{0.00026978683471679687} &  1.26 &   \num{0.0003774261474609375} &    0.81 &   \num{0.8519236845970154} \\
   &     & 0.9 &  2.22 &  \num{0.00030082559585571287} &  1.36 &  \num{0.00041724777221679695} &    1.61 &   \num{0.8919827847480775} \\
   & 0.6 & 0.1 &  3.28 &  \num{0.00025621747970581055} &  2.41 &  \num{0.00038323116302490227} &    2.29 &   \num{0.7056541695594788} \\
   &     & 0.4 &  1.93 &   \num{0.0002857007980346679} &  1.50 &  \num{0.00040060138702392576} &    1.32 &   \num{0.9110462455749513} \\
   &     & 0.9 &  1.78 &  \num{0.00027794837951660155} &  1.43 &   \num{0.0004000129699707032} &    1.23 &   \num{0.9203682947158812} \\
   & 0.9 & 0.1 &  4.17 &   \num{0.0002637534141540527} &  3.68 &   \num{0.0003872451782226563} &    3.72 &   \num{0.8124037542343139} \\
   &     & 0.4 &  2.32 &  \num{0.00033046340942382814} &  1.67 &  \num{0.00045054769515991204} &    1.69 &   \num{0.8606182594299318} \\
   &     & 0.6 &  2.24 &   \num{0.0002883424758911133} &  1.83 &  \num{0.00039455366134643553} &    1.57 &   \num{0.9533530807495116} \\
\hline
12 & 0.1 & 0.4 &  2.39 &    \num{0.000538719654083252} &  1.79 &   \num{0.0006898856163024902} &    1.30 &   \num{0.9466131477355957} \\
   &     & 0.6 &  4.82 &  \num{0.00044567918777465817} &  2.99 &   \num{0.0007670159339904784} &    3.06 &    \num{0.956727442741394} \\
   &     & 0.9 &  5.54 &   \num{0.0004638071060180664} &  3.81 &   \num{0.0007291016578674315} &    3.63 &    \num{1.018367877006531} \\
   & 0.4 & 0.1 &  2.30 &  \num{0.00039853429794311525} &  2.16 &   \num{0.0006596550941467285} &    1.69 &      \num{1.0639802069664} \\
   &     & 0.6 &  1.47 &   \num{0.0005176906585693359} &  1.14 &   \num{0.0007247509956359863} &    0.80 &    \num{1.355100109577179} \\
   &     & 0.9 &  1.95 &   \num{0.0004995026588439941} &  1.95 &   \num{0.0006955432891845705} &    1.83 &     \num{1.22169517993927} \\
   & 0.6 & 0.1 &  2.40 &  \num{0.00044720268249511723} &  2.08 &   \num{0.0007394895553588867} &    1.94 &   \num{1.0840875015258788} \\
   &     & 0.4 &  1.58 &   \num{0.0004239773750305175} &  1.30 &   \num{0.0006029324531555175} &    1.22 &    \num{1.228016318321228} \\
   &     & 0.9 &  1.75 &    \num{0.000603236198425293} &  1.57 &   \num{0.0007744293212890625} &    1.25 &   \num{1.2530505995750427} \\
   & 0.9 & 0.1 &  3.07 &   \num{0.0005225682258605958} &  2.68 &   \num{0.0007334089279174804} &    2.45 &   \num{1.0962561349868776} \\
   &     & 0.4 &  1.90 &   \num{0.0005604739189147949} &  1.62 &   \num{0.0008226404190063477} &    1.28 &   \num{1.1944667158126834} \\
   &     & 0.6 &  1.35 &   \num{0.0005491480827331544} &  1.19 &   \num{0.0007291064262390137} &    0.88 &   \num{1.2651404547691347} \\
\hline
14 & 0.1 & 0.4 &  2.59 &   \num{0.0007247629165649413} &  2.33 &   \num{0.0010779647827148437} &    1.49 &   \num{1.3647663106918335} \\
   &     & 0.6 &  3.50 &   \num{0.0007922196388244629} &  2.76 &   \num{0.0011821465492248536} &    1.95 &   \num{1.3897056016921998} \\
   &     & 0.9 &  5.61 &   \num{0.0008356504440307618} &  4.38 &    \num{0.001115994930267334} &    3.85 &   \num{1.6390694408416746} \\
   & 0.4 & 0.1 &  2.06 &   \num{0.0007080187797546387} &  1.81 &   \num{0.0009973626136779786} &    1.25 &   \num{1.5727275195121764} \\
   &     & 0.6 &  1.21 &   \num{0.0007431545257568361} &  1.15 &   \num{0.0009741554260253906} &    0.69 &   \num{1.6741085238456723} \\
   &     & 0.9 &  1.64 &   \num{0.0007624931335449218} &  1.43 &   \num{0.0011147165298461914} &    0.98 &   \num{2.0232930121421813} \\
   & 0.6 & 0.1 &  3.76 &   \num{0.0006480178833007811} &  3.86 &   \num{0.0010992794036865233} &    2.73 &    \num{4.990627601146699} \\
   &     & 0.4 &  1.23 &    \num{0.000784172534942627} &  1.06 &   \num{0.0010604362487792967} &    0.75 &    \num{6.160361298084259} \\
   &     & 0.9 &  1.39 &   \num{0.0008637876510620117} &  1.23 &   \num{0.0012183012962341307} &    0.94 &    \num{4.124634238243103} \\
   & 0.9 & 0.1 &  4.23 &   \num{0.0007876858711242674} &  3.73 &    \num{0.001179515838623047} &    3.12 &   \num{1.6282504434585572} \\
   &     & 0.4 &  1.48 &   \num{0.0008195786476135253} &  1.28 &   \num{0.0010700860023498536} &    0.98 &   \num{1.8188820428848267} \\
   &     & 0.6 &  1.61 &   \num{0.0008428015708923341} &  1.56 &   \num{0.0011696453094482422} &    1.24 &   \num{1.9321714911460877} \\
\hline
16 & 0.1 & 0.4 &  1.77 &   \num{0.0012574715614318847} &  1.54 &   \num{0.0017187809944152835} &    0.98 &   \num{2.0613828110694885} \\
   &     & 0.6 &  3.50 &   \num{0.0011214151382446288} &  3.09 &   \num{0.0017229962348937987} &    2.42 &   \num{2.0534218754768374} \\
   &     & 0.9 &  5.18 &   \num{0.0011065979003906249} &  4.16 &   \num{0.0018201527595520018} &    4.09 &   \num{2.0913944268226627} \\
   & 0.4 & 0.1 &  1.70 &   \num{0.0009863100051879883} &  1.59 &   \num{0.0014530901908874514} &    1.15 &    \num{7.588959093570708} \\
   &     & 0.6 &  1.15 &   \num{0.0011318750381469726} &  1.07 &   \num{0.0017002716064453127} &    0.45 &    \num{8.408233568668367} \\
   &     & 0.9 &  1.88 &   \num{0.0014100770950317382} &  1.83 &   \num{0.0018928813934326176} &    1.28 &    \num{8.150491786956788} \\
   & 0.6 & 0.1 &  2.52 &   \num{0.0010731434822082518} &  2.59 &   \num{0.0016813726425170897} &    2.04 &    \num{5.749234037876129} \\
   &     & 0.4 &  1.03 &    \num{0.001230994701385498} &  1.03 &   \num{0.0015921111106872558} &    0.52 &   \num{2.6797243838310245} \\
   &     & 0.9 &  1.16 &   \num{0.0011783790588378907} &  1.14 &   \num{0.0015985217094421386} &    0.71 &   \num{2.8341497645378113} \\
   & 0.9 & 0.1 &  3.78 &    \num{0.001081196308135986} &  4.03 &   \num{0.0016635389328002927} &    3.80 &   \num{2.1941118416786187} \\
   &     & 0.4 &  1.20 &   \num{0.0012204556465148925} &  1.15 &   \num{0.0017343573570251466} &    0.85 &    \num{2.611131365776062} \\
   &     & 0.6 &  0.94 &   \num{0.0011825976371765137} &  0.99 &   \num{0.0016957745552062988} &    0.59 &    \num{2.966841919898987} \\
\hline
   \multicolumn{3}{|c|}{Average} & 2.71 & \num{0.000573} & 2.07 & \num{0.000826} & 1.92 & \num{1.821161} \\
\hline
\end{tabular}
}
\caption{(Average) Relative percentage gap and solution times of the heuristic methods for S1}
\label{exp1-heuristics-avg-gap}
\end{table}

\begin{table}[htbp]
\centering
{\fontsize{9}{10.5}\selectfont
\begin{tabular}{|c|c|l|rr|rr|rr|}
\hline
& & Community & \multicolumn{2}{c|}{\texttt{E-LS1}} & \multicolumn{2}{c|}{\texttt{E-LS2}} & \multicolumn{2}{c|}{\texttt{E-exact}} \\
$K$ & $n$ & strength & Gap (\%) &     Time (s) &         Gap (\%) &     Time (s) &          Gap (\%) &     Time (s) \\
\hline
2 & 8  & low &  2.54 &  \num{0.00013491487503051758} &  1.69 &  \num{0.00017609834671020505} &    2.00 &  \num{1.5969632821083066} \\
  &    & medium &  3.14 &   \num{0.0001486649513244629} &  2.49 &  \num{0.00019037961959838867} &    2.32 &  \num{1.3477154984474182} \\
  &    & high &  5.81 &  \num{0.00014823055267333983} &  4.33 &  \num{0.00017769098281860353} &    5.82 &  \num{1.4627565131187439} \\
  & 10 & low &  2.45 &   \num{0.0002568554878234863} &  1.62 &   \num{0.0003673415184020996} &    1.81 &   \num{4.970001605033875} \\
  &    & medium &  2.40 &   \num{0.0002473244667053223} &  1.94 &   \num{0.0003467597961425781} &    1.85 &   \num{4.836548063755035} \\
  &    & high &  4.85 &   \num{0.0005157880783081055} &  3.63 &   \num{0.0004101457595825196} &    3.93 &   \num{4.822358114242554} \\
  & 12 & low &  1.47 &   \num{0.0004543204307556153} &  1.29 &   \num{0.0006091642379760743} &    0.91 &   \num{3.724962858200074} \\
  &    & medium &  3.09 &   \num{0.0004474015235900879} &  2.94 &   \num{0.0006405973434448242} &    2.33 &   \num{3.368069405555725} \\
  &    & high &  4.12 &  \num{0.00043682622909545904} &  3.56 &   \num{0.0006568474769592285} &    3.80 &   \num{3.340211583614349} \\
  & 14 & low &  2.18 &   \num{0.0007926220893859864} &  1.85 &   \num{0.0011294851303100587} &    2.65 &  \num{22.298229983806614} \\
  &    & medium &  2.21 &   \num{0.0008085727691650391} &  1.88 &   \num{0.0009812908172607423} &    2.00 &  \num{13.045899188518527} \\
  &    & high &  5.16 &   \num{0.0010807242393493653} &  4.33 &    \num{0.001192073345184326} &    4.64 &    \num{21.7898412733078} \\
  & 16 & low &  1.61 &    \num{0.001146953582763672} &  1.58 &   \num{0.0015468168258666991} &    1.11 &   \num{2.710313823699951} \\
  &    & medium &  2.27 &   \num{0.0012176318168640138} &  2.03 &   \num{0.0016551928520202636} &    1.60 &  \num{2.3824354634284974} \\
  &    & high &  4.51 &   \num{0.0011261358261108398} &  3.51 &     \num{0.00165169095993042} &    3.45 &  \num{2.7086530141830445} \\
\hline
3 & 8  & low &  5.35 &  \num{0.00048212957382202145} &  1.15 &   \num{0.0009660472869873045} &    4.55 &  \num{1.2869431653022767} \\
  &    & medium &  5.90 &  \num{0.00043556118011474603} &  1.64 &    \num{0.000916821002960205} &    5.52 &  \num{1.1573011059761047} \\
  &    & high &  5.95 &  \num{0.00044049549102783206} &  1.90 &   \num{0.0008237400054931642} &    5.75 &  \num{1.1363313231468202} \\
  & 10 & low &  3.81 &   \num{0.0007999601364135741} &  1.94 &   \num{0.0017158026695251465} &    3.50 &  \num{4.7856561336517345} \\
  &    & medium &  4.90 &   \num{0.0007559409141540527} &  2.01 &    \num{0.001800068855285645} &    4.28 &   \num{4.193556254863739} \\
  &    & high &  4.87 &   \num{0.0007115230560302734} &  2.35 &   \num{0.0016285719871520995} &    3.90 &   \num{4.214138635635375} \\
  & 12 & low &  3.25 &   \num{0.0016603512763977053} &  2.10 &   \num{0.0029364314079284672} &    2.68 &   \num{7.642225312232971} \\
  &    & medium &  3.98 &   \num{0.0017502942085266117} &  2.21 &    \num{0.003452012538909912} &    9.01 &   \num{18.16930490875244} \\
  &    & high &  4.55 &    \num{0.001794489860534668} &  2.43 &   \num{0.0032722029685974124} &    4.15 &   \num{6.129310718536378} \\
  & 14 & low &  2.59 &    \num{0.003159794807434082} &  1.69 &     \num{0.00434356689453125} &    2.03 &   \num{6.164107033729554} \\
  &    & medium &  2.67 &    \num{0.002879765033721923} &  1.46 &    \num{0.004821932792663574} &    2.38 &    \num{5.98454501247406} \\
  &    & high &  4.17 &   \num{0.0028682031631469727} &  1.96 &   \num{0.0048289461135864265} &    3.40 &   \num{4.452762880802155} \\
  & 16 & low &  0.84 &    \num{0.005182864665985107} & -0.05 &    \num{0.007153315067291259} &    0.43 &  \num{12.135712614536285} \\
  &    & medium &  1.36 &    \num{0.005504992961883545} &  0.49 &     \num{0.00756531286239624} &    1.19 &  \num{11.378996646881102} \\
  &    & high &  3.32 &    \num{0.005070438861846924} &  1.85 &    \num{0.007824700355529785} &    2.34 &   \num{9.725722141742704} \\
\hline
\multicolumn{3}{|c|}{Average} & 3.51 & \num{0.001415} & 2.12 & \num{0.002193} & 3.17 & \num{6.432052} \\
\hline
\end{tabular}
}
\caption{(Average) Relative percentage gap and solution times of the heuristic methods for S2}
\label{exp2-heuristics-avg-gap}
\end{table}

In terms of solution quality, \texttt{EM-LS2} achieves a lower gap on average, compared to \texttt{EM-LS1}.
The comparison with \texttt{EM-exact} leads to more contrasted observations: for data group S1, \texttt{EM-exact} achieved the lowest average gap, whereas for S2 it was outperformed by \texttt{EM-LS2}. For some settings, the average gap obtained by the heuristics is small or even negative (e.g., for S2, with $K=3$, $n=16$ and low community strength), meaning in the latter case that the heuristic objective value was better than the one found by the exact method (only possible when the exact method was unable to find the optimal solution within the time limit).

On several runs, we observed that the heuristics effectively found the optimal solutions (or high-quality solutions).
This insight directly derives from our ability to find optimal solutions with the proposed exact algorithms, as the heuristics by themselves cannot give such a performance certificate. It also remains an open question whether this behavior holds for larger instance, but such an analysis would require significant methodological advances to solve larger cases to proven optimality.

\subsection{Comparison to the ground truth}

We finally compare the model parameters found by the exact methods with the ground truth parameters used in the generation of each instance.
In this analysis, we calculate the agreement~$A(\hat{\mathbf{Z}}, \mathbf{Z}^*)$ between the community assignments $\hat{\mathbf{Z}}$ of the optimal solution of maximum likelihood and the ground truth communities $\mathbf{Z}^*$ of the model.
The agreement function $A(\cdot, \cdot)$ measures the maximum number of common elements between two vectors of community assignments, considering all possible permutations of the community labels.
When the optimum is not known, the estimated communities of the BKS are considered instead.

\begin{figure}[t]
\includegraphics[width=\textwidth]{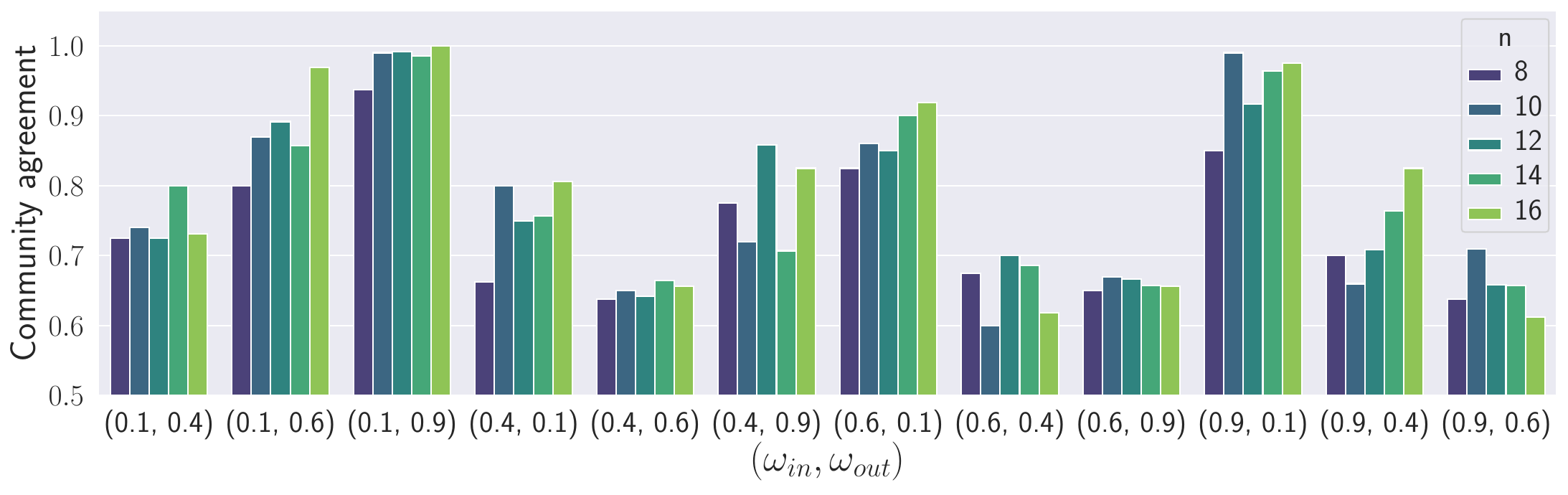}
\centering
\caption{(Average) agreement between the community assignments of maximum likelihood and the ground truth, as a function of $n$ and $(\omega_{\text{in}}, \omega_{\text{out}})$, for data sets in group S1}
\label{fig:exp1_community_agreements}
\end{figure}

\begin{figure}[t]
\includegraphics[width=\textwidth]{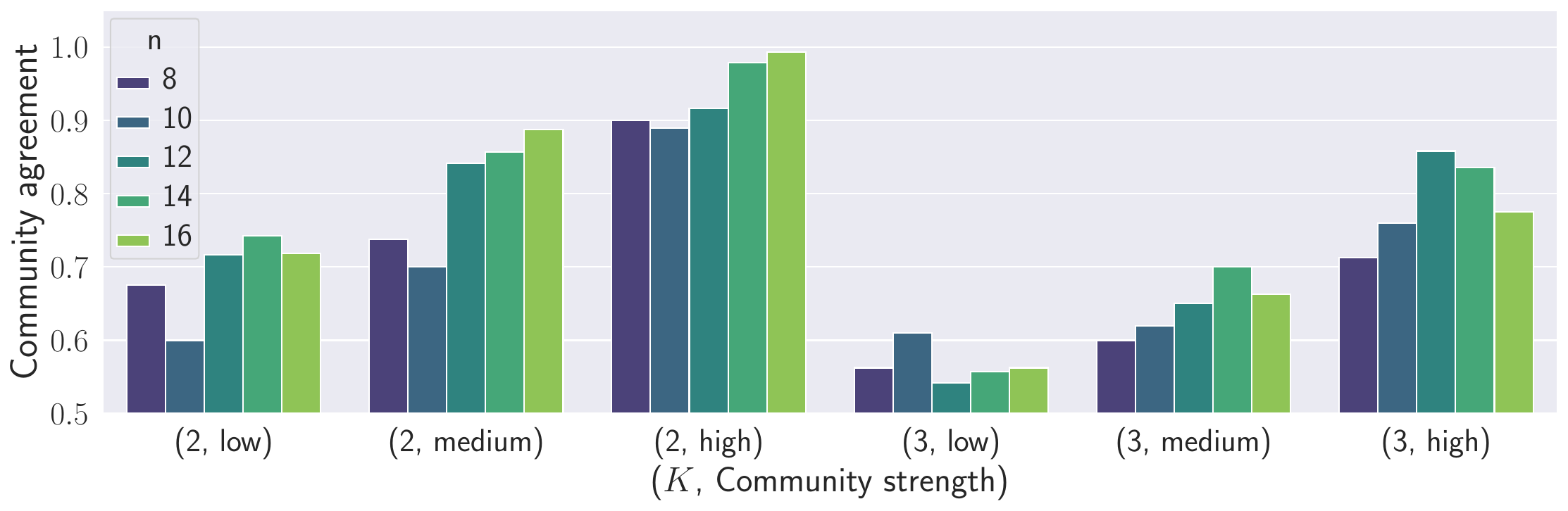}
\centering
\caption{(Average) agreement between the community assignments of maximum likelihood and the ground truth, as a function of $n$, $K$ and the level of community strength, for data sets in group S2}
\label{fig:exp2_community_agreements}
\end{figure}

Figure \ref{fig:exp1_community_agreements} shows that the agreement between the recovered communities and the ground truth communities of data set S1 is higher when $n$ is larger and when the absolute difference $|\omega_{\text{in}} - \omega_{\text{out}}|$ is larger. 
This is expected since there is more information in the graph.
Similarly for S2, we observe that instances with higher community strength have a higher community agreement (Figure \ref{fig:exp2_community_agreements}). 

The solution of maximum likelihood may be far from the ground truth, in general.
This is especially true for small networks, such as those considered in this work, since there is often not enough information to correctly recover the underlying communities.

We also compare the heuristics in their ability to recover the ground truth communities in Tables \ref{exp1-agreement-avg-heuristics} and \ref{exp2-agreement-avg-heuristics}, respectively for S1 and S2.
These tables report the average agreement (out of 50 trials) between the recovered communities and the ground truth communities used in the generation of each instance. 
We compare the performance of the heuristics with the MILP with SBC.

\begin{table}[htbp]
\centering
\vspace*{-1.25cm}
{\fontsize{9}{11}\selectfont
\begin{tabular}{|c|cc|cccc|}
\hline
 & & & \multicolumn{4}{c|}{Average agreement}\\
$n$ & $\omega_{\text{in}}$ & $\omega_{\text{out}}$ & \texttt{E-LS1} &  \texttt{E-LS2} & \texttt{E-exact} & MILP \\
\hline
8  & 0.1 & 0.4 &           0.69 &  \textbf{0.73} &           0.71 &           0.72 \\
   &     & 0.6 &           0.72 &           0.75 &           0.74 &   \textbf{0.8} \\
   &     & 0.9 &           0.76 &           0.85 &           0.82 &  \textbf{0.94} \\
   & 0.4 & 0.1 &           0.65 &  \textbf{0.67} &           0.64 &           0.66 \\
   &     & 0.6 &           0.66 &           0.66 &  \textbf{0.69} &           0.64 \\
   &     & 0.9 &           0.73 &           0.75 &           0.75 &  \textbf{0.78} \\
   & 0.6 & 0.1 &           0.68 &           0.70 &           0.68 &  \textbf{0.82} \\
   &     & 0.4 &           0.66 &           0.67 &           0.65 &  \textbf{0.68} \\
   &     & 0.9 &           0.63 &           0.64 &           0.62 &  \textbf{0.65} \\
   & 0.9 & 0.1 &           0.72 &           0.78 &           0.75 &  \textbf{0.85} \\
   &     & 0.4 &           0.66 &           0.67 &           0.68 &   \textbf{0.7} \\
   &     & 0.6 &           0.63 &  \textbf{0.64} &           0.63 &  \textbf{0.64} \\
\hline
10 & 0.1 & 0.4 &           0.67 &           0.71 &           0.72 &  \textbf{0.74} \\
   &     & 0.6 &           0.74 &           0.78 &           0.77 &  \textbf{0.87} \\
   &     & 0.9 &           0.78 &           0.81 &           0.85 &  \textbf{0.99} \\
   & 0.4 & 0.1 &           0.68 &           0.69 &           0.71 &   \textbf{0.8} \\
   &     & 0.6 &  \textbf{0.66} &           0.65 &           0.65 &           0.65 \\
   &     & 0.9 &           0.68 &           0.68 &  \textbf{0.73} &           0.72 \\
   & 0.6 & 0.1 &           0.72 &           0.75 &           0.75 &  \textbf{0.86} \\
   &     & 0.4 &           0.62 &  \textbf{0.63} &           0.62 &           0.60 \\
   &     & 0.9 &           0.63 &           0.64 &           0.63 &  \textbf{0.67} \\
   & 0.9 & 0.1 &           0.79 &           0.79 &           0.80 &  \textbf{0.99} \\
   &     & 0.4 &  \textbf{0.66} &  \textbf{0.66} &  \textbf{0.66} &  \textbf{0.66} \\
   &     & 0.6 &           0.65 &           0.65 &           0.65 &  \textbf{0.71} \\
\hline
12 & 0.1 & 0.4 &           0.67 &           0.69 &           0.70 &  \textbf{0.72} \\
   &     & 0.6 &           0.75 &           0.80 &           0.81 &  \textbf{0.89} \\
   &     & 0.9 &           0.76 &           0.81 &           0.82 &  \textbf{0.99} \\
   & 0.4 & 0.1 &           0.67 &           0.68 &           0.70 &  \textbf{0.75} \\
   &     & 0.6 &           0.63 &           0.62 &  \textbf{0.65} &           0.64 \\
   &     & 0.9 &           0.72 &           0.70 &           0.71 &  \textbf{0.86} \\
   & 0.6 & 0.1 &           0.71 &           0.71 &           0.74 &  \textbf{0.85} \\
   &     & 0.4 &           0.66 &           0.68 &           0.67 &   \textbf{0.7} \\
   &     & 0.9 &           0.64 &           0.63 &           0.66 &  \textbf{0.67} \\
   & 0.9 & 0.1 &           0.75 &           0.77 &           0.78 &  \textbf{0.92} \\
   &     & 0.4 &           0.67 &           0.66 &  \textbf{0.71} &  \textbf{0.71} \\
   &     & 0.6 &           0.64 &           0.65 &           0.63 &  \textbf{0.66} \\
\hline
14 & 0.1 & 0.4 &           0.69 &           0.69 &           0.73 &   \textbf{0.8} \\
   &     & 0.6 &           0.70 &           0.73 &           0.76 &  \textbf{0.86} \\
   &     & 0.9 &           0.78 &           0.81 &           0.84 &  \textbf{0.99} \\
   & 0.4 & 0.1 &           0.67 &           0.69 &           0.67 &  \textbf{0.76} \\
   &     & 0.6 &           0.63 &           0.62 &           0.64 &  \textbf{0.66} \\
   &     & 0.9 &           0.65 &           0.65 &           0.67 &  \textbf{0.71} \\
   & 0.6 & 0.1 &           0.74 &           0.72 &           0.74 &   \textbf{0.9} \\
   &     & 0.4 &           0.63 &           0.62 &           0.65 &  \textbf{0.69} \\
   &     & 0.9 &           0.61 &           0.62 &           0.61 &  \textbf{0.66} \\
   & 0.9 & 0.1 &           0.75 &           0.77 &           0.79 &  \textbf{0.96} \\
   &     & 0.4 &           0.66 &           0.66 &           0.67 &  \textbf{0.76} \\
   &     & 0.6 &           0.63 &           0.63 &           0.62 &  \textbf{0.66} \\
\hline
16 & 0.1 & 0.4 &           0.65 &           0.66 &           0.71 &  \textbf{0.73} \\
   &     & 0.6 &           0.77 &           0.78 &           0.81 &  \textbf{0.97} \\
   &     & 0.9 &           0.80 &           0.82 &           0.84 &   \textbf{1.0} \\
   & 0.4 & 0.1 &           0.68 &           0.68 &           0.70 &  \textbf{0.81} \\
   &     & 0.6 &           0.62 &           0.62 &           0.63 &  \textbf{0.66} \\
   &     & 0.9 &           0.70 &           0.69 &           0.75 &  \textbf{0.82} \\
   & 0.6 & 0.1 &           0.75 &           0.74 &           0.77 &  \textbf{0.92} \\
   &     & 0.4 &           0.61 &  \textbf{0.62} &           0.61 &  \textbf{0.62} \\
   &     & 0.9 &           0.65 &           0.64 &           0.65 &  \textbf{0.66} \\
   & 0.9 & 0.1 &           0.79 &           0.79 &           0.79 &  \textbf{0.98} \\
   &     & 0.4 &           0.69 &           0.70 &           0.77 &  \textbf{0.82} \\
   &     & 0.6 &  \textbf{0.61} &  \textbf{0.61} &  \textbf{0.61} &  \textbf{0.61} \\
\hline
   \multicolumn{3}{|c|}{Average} & 0.69 & 0.70 & 0.71 & \textbf{0.77} \\
\hline
\end{tabular}
}
\caption{Comparison between heuristic and exact solution algorithms in recovering the ground truth communities of data sets in group S1}
\label{exp1-agreement-avg-heuristics}
\end{table}

\begin{table}[htbp]
\centering
\setlength{\tabcolsep}{4pt}
{\fontsize{9}{11}\selectfont
\begin{tabular}{|c|c|l|cccc|}
\hline
 & & Community &  \multicolumn{4}{c|}{Average agreement}\\
$K$ & $n$ & strength & \texttt{E-LS1} &  \texttt{E-LS2} & \texttt{E-exact} & MILP \\
\hline
2 & 8  & low &           0.64 &           0.65 &     0.63 &  \textbf{0.68} \\
  &    & medium &           0.67 &           0.67 &     0.67 &  \textbf{0.74} \\
  &    & high &           0.75 &           0.77 &     0.74 &   \textbf{0.9} \\
  & 10 & low &  \textbf{0.63} &           0.62 &     0.60 &           0.60 \\
  &    & medium &           0.68 &           0.68 &     0.64 &   \textbf{0.7} \\
  &    & high &           0.75 &           0.77 &     0.78 &  \textbf{0.89} \\
  & 12 & low &           0.65 &           0.65 &     0.67 &  \textbf{0.72} \\
  &    & medium &           0.70 &           0.72 &     0.71 &  \textbf{0.84} \\
  &    & high &           0.80 &           0.82 &     0.82 &  \textbf{0.92} \\
  & 14 & low &           0.68 &           0.69 &     0.67 &  \textbf{0.74} \\
  &    & medium &           0.70 &           0.71 &     0.70 &  \textbf{0.86} \\
  &    & high &           0.76 &           0.79 &     0.79 &  \textbf{0.98} \\
  & 16 & low &           0.66 &           0.67 &     0.69 &  \textbf{0.72} \\
  &    & medium &           0.76 &           0.76 &     0.80 &  \textbf{0.89} \\
  &    & high &           0.77 &           0.82 &     0.82 &  \textbf{0.99} \\
\hline
3 & 8  & low &           0.58 &  \textbf{0.59} &     0.57 &           0.56 \\
  &    & medium &           0.57 &   \textbf{0.6} &     0.56 &   \textbf{0.6} \\
  &    & high &           0.63 &           0.69 &     0.65 &  \textbf{0.71} \\
  & 10 & low &           0.56 &           0.56 &     0.57 &  \textbf{0.61} \\
  &    & medium &           0.60 &           0.60 &     0.61 &  \textbf{0.62} \\
  &    & high &           0.61 &           0.66 &     0.61 &  \textbf{0.76} \\
  & 12 & low &  \textbf{0.56} &  \textbf{0.56} &     0.54 &           0.54 \\
  &    & medium &           0.62 &           0.62 &     0.57 &  \textbf{0.65} \\
  &    & high &           0.66 &           0.72 &     0.67 &  \textbf{0.86} \\
  & 14 & low &           0.54 &           0.54 &     0.54 &  \textbf{0.56} \\
  &    & medium &           0.60 &           0.63 &     0.61 &   \textbf{0.7} \\
  &    & high &           0.68 &           0.73 &     0.70 &  \textbf{0.84} \\
  & 16 & low &           0.53 &           0.54 &     0.54 &  \textbf{0.56} \\
  &    & medium &           0.61 &           0.62 &     0.63 &  \textbf{0.66} \\
  &    & high &           0.66 &           0.69 &     0.69 &  \textbf{0.78} \\
\hline
\multicolumn{3}{|c|}{Average} & 0.65 & 0.67 & 0.66 & \textbf{0.74} \\
\hline
\end{tabular}
}
\caption{Comparison between heuristic and exact solution algorithms in recovering the ground truth communities of data sets in group S2}
\label{exp2-agreement-avg-heuristics}
\end{table}

For some instances, the resulting community agreement is low for both exact and heuristic methods, since there is not enough information present in the graph and it may be theoretically impossible to recover the ground truth. Still, in the other cases, the exact approach clearly outperforms the heuristics in almost all instances, highlighting the importance of good solutions for this task.

\section{Conclusions}
\label{section-conclusions}

This study allowed us to fill a significant methodological gap: the lack of exact solution methods for community detection in the general SBM. Exact algorithms are indeed essential for a disciplined analysis of machine learning models and training algorithms, as they permit a precise evaluation of heuristic performance. The goal of a heuristic is to achieve an optimality gap that is systematically close to 0\% for the model at hand. As heuristics do not provide guarantees regarding solution quality, we cannot evaluate their true optimality gap unless we have access to an efficient algorithm that produces optimal solutions (or at least good bounds on solution value).

To that end, we have introduced new mathematical programming formulations for the MLE model of the DCSBM. We introduced a descriptive formulation based on a MINLP and employed linearization techniques to transform it into a MILP. We also proposed bound tightening and symmetry-breaking strategies, which lead to critical improvements to the model. The proposed solution methods can find optimal solutions of maximum likelihood with a certificate of global optimality. Furthermore, we have reviewed three natural variants of the EM algorithm for this problem, and conducted extensive numerical analyses to analyze their performance.

This work raises several interesting avenues for future research.
In particular, there is still space to improve the scalability of the exact methods. In our computational experiments, we noted that the MILP often identifies the optimal solution early in the optimization process but that it takes a much longer time to find good lower bounds and prove optimality. To improve this behavior, research could be pursued on new problem formulations and valid inequalities permitting to achieve tighter lower bounds and enhance the efficiency of the branch-and-bound exploration. Another alternative is to explore mathematical decomposition techniques such as column generation, which have the potential to lead to structurally-different formulations and solution approaches. Finally, we generally encourage the pursuit of a disciplined analysis of algorithms for other learning tasks, and likewise develop mathematical programming approaches for other models of importance.

\section*{Acknowledgements}

This research has been partially funded by CAPES, CNPq [grant number 308528/2018-2] and FAPERJ [grant number E-26/202.790/2019] in Brazil, and by the Deutsche Forschungsgemeinschaft (DFG, German Research Foundation) [grant number 277991500/GRK2201] in Germany. This support is gratefully acknowledged.


\end{document}